\documentclass[12pt,onecolumn,draftclsnofoot]{IEEEtran}
\usepackage{xspace,amsmath,amssymb,amsfonts,epsfig,subfigure,syntonly}
\usepackage{lipsum}
\usepackage{cite,bm,color,url,textcomp,empheq,boxedminipage}
\usepackage{algorithmicx,algorithm}
\usepackage{indentfirst}
\usepackage{epstopdf,makecell}
\usepackage{empheq}
\usepackage{pifont}
\usepackage{stfloats}
\usepackage[noend]{algpseudocode}

\usepackage{graphicx,graphics}  
\usepackage{multirow,multicol}
\usepackage{psfrag}    
\usepackage{stfloats}
\usepackage{url}
\usepackage{lipsum}
\usepackage{blindtext}

\newcommand{\mv}[1]{\mbox{\boldmath{$ #1 $}}}

\usepackage{amssymb}
\usepackage{amsmath}
\usepackage{cite}
\usepackage{url}
\usepackage{xcolor}
\usepackage{cite,graphicx,amsmath,amssymb}
\usepackage{subfigure}
\usepackage{fancyhdr}
\usepackage{mdwmath}
\usepackage{mdwtab}
\usepackage{caption}
\usepackage{amsthm}
\usepackage{setspace}

\graphicspath{{figures/}}

\begin{document}

\title{Channel Knowledge Map (CKM)-Assisted Multi-UAV Wireless Network: CKM Construction and UAV Placement}

\author{

Haoyun~Li,~
        Peiming~Li,~
        Jie~Xu,~
        Junting~Chen,~
        and Yong~Zeng
\thanks{Part of this paper has been presented at the IEEE International Conference on Communications (ICC), Seoul, South Korea, 2022 \cite{conf}.}
\thanks{H. Li, J. Xu, P. Li, and J. Chen are with the School of Science and Engineering, and the Future Network of Intelligence Institute (FNii), The Chinese University of Hong Kong, Shenzhen, Shenzhen 518172, China (e-mail: haoyunli@link.cuhk.edu.cn, \{xujie, juntingc\}@cuhk.edu.cn, peiminglee@outlook.com). J. Xu is the corresponding author.}
\thanks{Y. Zeng is with the National Mobile Communications Research
Laboratory and Frontiers Science Center for Mobile Information Communication and Security, Southeast University, Nanjing 210096, China. He is also
with the Purple Mountain Laboratories, Nanjing 211111, China (e-mail: yong\_zeng@seu.edu.cn).}
}
\maketitle
\begin{abstract}
Channel knowledge map (CKM) has recently emerged as a viable new solution to facilitate the placement and trajectory optimization for unmanned aerial vehicle (UAV) communications, by exploiting the site- and location-specific radio propagation information. This paper investigates a CKM-assisted multi-UAV wireless network, by focusing on the construction and utilization of CKMs for multi-UAV placement optimization. First, we consider the CKM construction problem when data measurements for only a limited number of points are available. Towards this end, we exploit a data-driven interpolation technique, namely the Kriging method, to construct CKMs to characterize the signal propagation environments. Next, we study the multi-UAV placement optimization problem by utilizing the constructed CKMs, in which the multiple UAVs aim to optimize their placement locations to maximize the weighted sum rate with their respectively associated ground base stations (GBSs). However, the weighted sum rate function based on the CKMs is generally non-differentiable, which renders the conventional optimization techniques relying on function derivatives inapplicable. To tackle this issue, we propose a novel iterative algorithm based on derivative-free optimization, in which a series of quadratic functions are iteratively constructed to approximate the objective function under a set of interpolation conditions, and accordingly, the UAVs' placement locations are updated by maximizing the approximate function subject to a trust region constraint. Finally, numerical results are presented to validate the performance of the proposed designs. It is shown that the Kriging method can construct accurate CKMs for UAVs. Furthermore, the proposed derivative-free placement optimization design based on the Kriging-constructed CKMs achieves a weighted sum rate that is close to the optimal exhaustive search design based on ground-truth CKMs, but with much lower implementation complexity. In addition, the proposed design is shown to significantly outperform other benchmark schemes.
\end{abstract}
\section{Introduction}
Unmanned aerial vehicle (UAV) communications have emerged as one of the key components for beyond the fifth-generation (B5G) and the sixth-generation (6G) wireless networks, in which UAVs can either act as aerial communication users with their own flying tasks (e.g., aerial package delivery), or serve as aerial platforms (such as base stations (BSs) and relays) to provide communication services to terrestrial subscribers in emerging scenarios and hot-spot areas \cite{UAV0,UAVtuto,UAV3,UAV6}. In contrast to conventional terrestrial communications, UAVs at a relatively high altitude have a high probability to experience line-of-sight (LoS) air-ground channels. Due to the strong air-ground channels, UAVs as transmitters may cause severe co-channel interference to coexisting terrestrial BSs or users, and as receivers, may also suffer from severe interference from adjacent UAV BSs or users sharing the same spectrum \cite{UAV0}. Therefore, how to optimize the UAV communication performance while mitigating the air-ground interference is a challenging task to deal with in UAV wireless networks.



Thanks to the highly controllable mobility and agility, optimizing the UAV placement or trajectory in three-dimensional (3D) space yields a new design degree of freedom for enhancing the communication performance for both UAV users and UAV BSs (see, e.g., \cite{UAV0,ZYInitial,Chen1,Chen2,QQ,lipeiming2,QQ2,lixinmin,Valiulahi,lipeiming}). In particular, UAVs can be positioned close to intended transceivers but away from the unintended ones, thus enhancing the desired communication channel quality while suppressing harmful co-channel interference. There have been a large body of prior works that investigated the UAV placement and trajectory optimization under different single-UAV setups, such as UAV-enabled relay channels \cite{ZYInitial,Chen1,Chen2}, broadcast channels \cite{QQ}, and multiple access channels \cite{lipeiming2,CYou2}. There have also been other prior works studying the multi-UAV placement \cite{lixinmin,Valiulahi} and trajectory \cite{QQ2} optimization to mitigate the inter-UAV interference and enhance the communication performance. For ease of exposition, the majority of these prior works mainly assumed LoS, probabilistic LoS, or Rician fading air-ground channel models, based on which the UAV placement or trajectory can be designed by using conventional convex and non-convex optimization techniques. However, these simplified or statistical air-ground channel models only describe wireless channels in an average sense, and cannot capture the site- or location-specific radio propagation environments, e.g., due to the blockage and shadowing caused by buildings and vegetation. As a result, these over-simplified channel assumptions may lead to degraded UAV communication performance in practice.

To address the above issue, channel knowledge map (CKM) \cite{CKM, R1, R2} or radio map \cite{SZhang,SBi, RLevie, LWJ} has emerged as a viable new solution, which provides a site-specific database that contains location-specific channel-related information (e.g., channel power gains, shadowing, interference, and AoA/AoD \cite{CKM}) for enabling environment-aware wireless communications. Recently, CKM has found abundant applications in, e.g., training-free beam alignment for millimeter wave systems \cite{R1}, beam selection for reconfigurable intelligent surface (RIS) \cite{R2}, and localization and sensing \cite{RLevie, LWJ}. In addition, there have been a number of prior works (e.g., \cite{mo,juntingc,YHuang,ZY2}) investigating the CKM-assisted UAV communications in the scenario with one single UAV, in which the UAV trajectory or placement is designed based on the CKM for optimizing the UAV communication performance.

This paper studies a CKM-assisted multi-UAV wireless network, in which multiple UAV users flying at a fixed altitude send individual messages to their associated ground BSs (GBSs) over the same frequency band. The considered multi-UAV wireless network faces two challenges for the construction and utilization of CKMs. On one hand, it is difficult to accurately construct site-specific CKMs for different GBSs, as only limited channel measurement data can be obtained at specific locations due to practical constraints (e.g., constraints on the area topology and device energy limitation). On the other hand, it is also difficult to efficiently optimize the UAV trajectory/placement based on CKMs, as the CKMs contain complicated discrete location-specific channel knowledge without analytic model expressions in general, thus making the conventional optimization techniques inapplicable. This thus motivates our work in this paper to develop new design approaches to construct and utilize CKMs for multi-UAV wireless networks.

%
%
%
%
%
%

\subsection{Related Works}

Existing works on CKM construction can be generally classified into four categories, namely the model-based, ray tracing-based, machine learning, and interpolation methods. First, model-based CKM construction method \cite{CYou,AAi,MMo} aims to construct segmented air-ground channel models for UAV communications, in which different pre-determined model parameters are adopted at each segment. Nonetheless, although the model-based segmented CKM is easy to be used for theoretical analysis, it may be far away from the ground-truth channels, especially when the number of segments is limited. Next, the ray tracing-based CKM construction method \cite{KRizk, NSuga} aims to generate a large number of rays to model the wave propagation and penetration in the 3D environment, based on which the channel knowledge from the transmitter to different receiver locations can be acquired. However, the ray tracing process is time-consuming and computation-demanding (especially for outdoor scenarios with UAVs), and its accuracy highly depends on the accuracy of the 3D environment model.
 
Different from model-based and ray tracing-based methods, machine learning and interpolation methods belong to data-driven CKM construction approaches. On one hand, machine learning methods aim to find a direct mapping from any location to its output channel knowledge, while treating the explicit functional relationship as a blackbox. Example machine learning methods include expectation maximization (EM) \cite{EM}, maximum likelihood estimation (MLE) \cite{MLE}, deep Gaussian process (DGP) \cite{DGP}, and deep learning \cite{RLevie,UMasood,YTeganya,KSuto}. However, the machine learning-based CKM construction has demanding requirements on the size and quality of datasets, as well as high computation power for training machine learning models, which are very time-consuming and not easy to implement for real-time scenarios. On the other hand, the interpolation method aims to estimate channel knowledge at unknown locations based on a limited number of sample observation data within an area. Example interpolation methods include inverse distance weighting (IDW) \cite{IDW}, K-Nearest Neighbors (KNN) \cite{knn}, and Kriging \cite{kriging3}. Among the four CKM construction methods, the interpolation method is most promising for practical implementation, since it can construct accurate CKMs with low implementation complexity, especially when only a limited number of sampled measurements are available. However, to our best knowledge, how to use the interpolation methods to construct CKM for air-ground channels for UAV communications has not been investigated yet. 

In the literature, there are also a handful of existing works studying the UAV placement \cite{mo} and trajectory design \cite{SZhang,YHuang,ZY2,juntingc} based on CKMs for single-UAV communication scenarios. In these works, the authors utilized techniques like graph theory \cite{SZhang}, exhaustive search \cite{mo}, and reinforcement learning \cite{YHuang,ZY2}, or simplified the general CKM as segmented channel models \cite{juntingc}. However, these designs are inapplicable for our considered multi-UAV wireless network with general CKMs, in which the channel knowledge from such CKMs cannot be characterized by any analytic functions with respect to the locations of transceivers. Therefore, new design techniques for placement optimization in multi-UAV wireless networks are needed.

\subsection{Contributions}
\label{sec:intro-contri}
This paper studies the CKM construction and utilization for UAV placement optimization in a multi-UAV wireless network. The main contributions of this work are summarized as follows. 
\begin{itemize}
    \item First, we consider the construction of site-specific air-ground CKMs for the multi-UAV wireless network based on a finite number of sampled measurement data points. We present the Kriging-based spatial interpolation algorithm to construct the CKMs, in which a linear interpolation model is adopted. In particular, we minimize the variance of the estimation error between the sampled measurement points and the corresponding estimation to obtain the weighting factors of linear interpolation.
%
%
%

    \item Next, based on the constructed CKMs, we optimize the placement locations of multiple UAVs to maximize their weighted sum rate. As CKMs normally contain discrete location-specific channel knowledge without analytic model functions, the corresponding weighted sum rate function becomes non-differentiable, thus making the conventional convex or non-convex optimization methods inapplicable. To tackle this issue, we adopt a novel method called derivative-free optimization \cite{toolbook}, which iteratively constructs a series of quadratic functions to approximate the objective function under a set of interpolation conditions, and accordingly updates the optimization variables by maximizing the approximate function subject to a trust region constraint. The convergence of the proposed algorithm can be ensured by properly designing the trust region.
    \item Finally, numerical results are presented to validate the performance of the proposed designs. It is shown that the constructed CKM based on Kriging method is close to the ground-truth values, and achieves lower construction error than the benchmarking KNN-based method. It is also shown that the proposed derivative-free placement optimization based on Kriging-estimated CKMs achieves a weighted sum rate close to the optimal value obtained by exhaustive search based on ground-truth CKMs, but with much lower implementation complexity. Furthermore, the proposed design is shown to significantly outperform the conventional optimization method based on simplified LoS channel models and the heuristic design with each UAV hovering above its associated GBS.
\end{itemize}
\subsection{Organization}
\label{sec:intro-orga}
The remainder of this paper is organized as follows. Section \ref{sec:sys} presents the system model. Section \ref{sec2} discusses the construction of CKMs using the Kriging algorithm. Section \ref{sec3} develops the derivative-free optimization to design the multi-UAV placement based on CKM. Section \ref{sec4} provides numerical results to validate the performance of Kriging-based CKM construction and derivative-free placement optimization. Section \ref{sec5} finally concludes this paper.
\section{System Model}
\label{sec:sys}
\begin{figure}[t]
        \centering
        \includegraphics[scale=0.8]{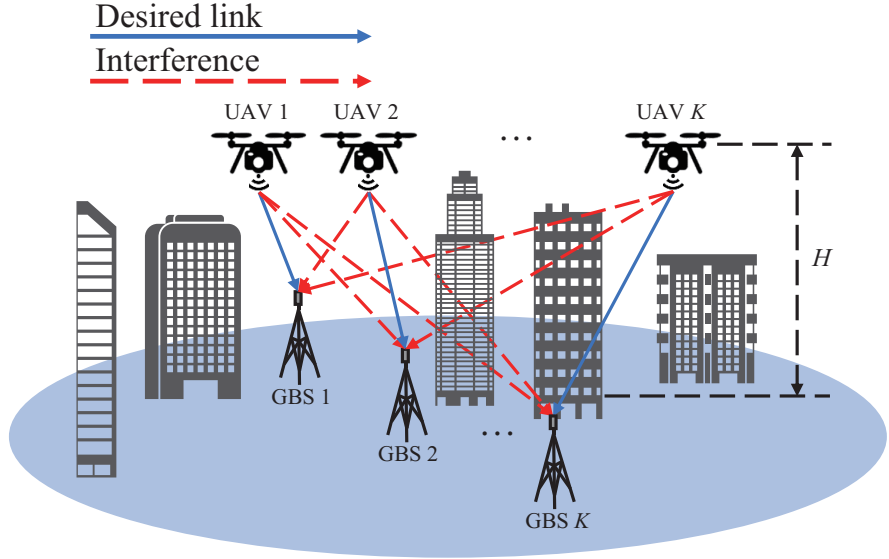}
        \caption{Illustration of the multi-UAV wireless network.}
        \label{system}
\end{figure}

We consider a multi-UAV wireless network as shown in Fig. \ref{system}, in which $K$ UAV users send individual messages to their respectively associated GBSs over the same frequency band. Let ${\cal K}\triangleq\{1,..., K\}$ denote the set of UAV users or GBSs. Each GBS $k \in {\cal K}$ is located at fixed location $(\hat{x}_k, \hat{y}_k, \hat{H})$ in a 3D coordinate system, where $\hat{H} \geq 0$ in meters (m) denotes the GBSs' height, and $\mathbf{w}_{k} = (\hat{x}_k, \hat{y}_k)$ denotes the horizontal location. Let $(x_j, y_j, H)$ denote the location of UAV $j\in {\cal K}$, where $\mathbf{q}_{j} = (x_j, y_j)$ denotes the horizontal location of UAV $j$ to be optimized, and $H$ denotes the altitude of the UAV that is assumed to be fixed. 

\subsection{CKM}

Site-specific CKMs are employed to store the location-specific wireless channel knowledge related to the GBSs. In particular, we assume that there is one site-specific CKM $\mathcal G_k$ for each GBS $k\in \cal K$, which provides mapping from any given horizontal location $\mathbf{q}$ at the constant altitude $H$ to the corresponding channel power gain\footnote{The channel power gain refers to the large-scale channel gain including path loss and shadowing, since the small-scale fading is hard to be obtained.} $Z_k$, i.e., $\mathbf{q}\in{\mathbb{R}}^{1\times2} \rightarrow Z_k\in{\mathbb{R}}$. Notice that the CKM should be practically constructed by, e.g., interpolation techniques based on the measurement data (as will be discussed in Section \ref{sec2}). Therefore, the channel power gain stored in the constructed CKM $\mathcal G_k$ may be slightly different from the ground-truth values.

For ease of exposition, we denote $\tilde{\mathcal G}_k$ as the ground-truth CKM for GBS $k\in \cal K$, which provides mapping from horizontal location $\mathbf{q}$ at altitude $H$ to the corresponding ground-truth channel power gain $\tilde Z_k$, i.e., $\mathbf{q}\in{\mathbb{R}}^{1\times2} \rightarrow \tilde Z_k\in{\mathbb{R}}$. Consider a particular GBS $k\in \mathcal K$. Suppose that there are $N_k$ pairs of measurement samples $\{(\mathbf{x}_{i}^{(k)}, \tilde Z_k(\mathbf{x}_{i}^{(k)})), i=1, \dots ,N_k\}$ with $\tilde Z_k(\mathbf{x}_{i}^{(k)}) = \tilde{\mathcal G}_k(\mathbf{x}_{i}^{(k)})$, which are collected by a UAV flying at fixed altitude $H$ within a specific area $\mathcal{D} \in \mathbb{R}^{1\times2}$. $\tilde Z_k(\mathbf{x})$ follows a second-order stationary process and is isotropic. Here, $\mathbf{x}_{i}^{(k)} \in \mathcal{D}$ denotes the location of the sample point, and $\tilde Z_k(\mathbf{x}_{i}^{(k)})$ denotes the ground-truth channel power gain at $\mathbf{x}_{i}^{(k)}$.

The CKM $\mathcal G_k$ for each GBS $k\in\mathcal K$ will be constructed based on the measurement points $\{(\mathbf{x}_{i}^{(k)}, \tilde Z_k(\mathbf{x}_{i}^{(k)})), i=1, \dots ,N_k\}$ in Section \ref{sec3}, which works as a site-specific database containing discrete location-specific channel knowledge, and will then be used to facilitate the UAV placement optimization in Section \ref{sec4}.

%

\subsection{Multi-UAV Communications}

The multiple UAVs design their placement locations based on the CKM $\{\mathcal G_k\}$. Based on the CKM, the predicted channel power gain between each UAV $j \in \cal K $ and GBS $k\in \cal K$ is given by
\begin{equation}
    \label{CKM:gain}
    Z_{k}(\mathbf{q}_{j})=\mathcal G_{k}(\mathbf{q}_{j}),
\end{equation}which is a discrete function without any analytic expression in general. 
Based on the channel power gain in \eqref{CKM:gain}, the received signal-to-interference-plus-noise ratio (SINR) at GBS $k\in{\cal K}$ is given by
\begin{equation}
    \gamma_{k}(\{\mathbf{q}_{j}\})=\frac{P_{k} Z_{k}(\mathbf{q}_{k})}{{\sum\nolimits_{j \in {\cal K}, j \neq {k}}P_{j} Z_{k}(\mathbf{q}_{j})} +\sigma^{2}_k},
\end{equation}where $P_{k}$ denotes the transmit power of UAV $k$, and $\sigma^{2}_k$ denotes the noise power at the receiver of GBS $k$, $k\in{\cal K}$. In order to focus our study on the placement optimization, we assume that the transmit power $P_k$'s at different UAVs are given, and leave the optimization of $P_k$'s in future work. By considering Gaussian signalling at each UAV transmitter, the achievable data rate from UAV $k \in{\cal K}$ to its associated GBS $k$ in bits/second/Hertz (bps/Hz) is given by
\begin{equation}
    r_{k}(\{\mathbf{q}_{j}\})=\log _{2}(1+\gamma_{k}(\{\mathbf{q}_{j}\})).
\end{equation}

Our objective is to maximize the weighted sum rate of UAVs. Let $\mv \alpha=\{\alpha_1,...,\alpha_K\}$ denote the predetermined weights that
specify the communication rate priority among the $K$ UAVs with $\alpha_k > 0, \forall k \in \cal K$. Then, the weighted sum rate maximization is formulated as\footnote{Notice that in the UAV placement optimization problem (P1), the channel power gains are evaluated based on constructed CKM $\mathcal G_k$'s. Nevertheless, after determining the optimized UAV locations, we will use the ground-truth CKM $\tilde{\mathcal G}_k$'s to evaluate the achievable rate performance (see Section \ref{sec4}).}
\begin{align}
    \text{(P1)}: &\max\limits_{\{\mathbf{q}_{k}\}}~{\sum\nolimits_{k\in{\cal K}}\alpha_{k}r_{k}(\{\mathbf{q}_{j}\})}\notag\\
    &~~\text {s.t.}~\mathbf{q}_{k} \in \mathcal{D},\forall k \in {\cal K},\label{range}
\end{align}
where $\mathcal{D}$ denotes the target area that limits the locations of the UAVs. Notice that with the environment-aware channel gain offered by CKM, the objective function in problem (P1) is non-differentiable with respect to the UAV locations $\{\mathbf{q}_k\}$. In fact, it does not even have an explicit expression with respect to the UAVs' placement locations. Therefore, problem (P1) is very difficult to be solved, since the conventional convex/non-convex optimization methods are inapplicable. We will address problem (P1) in Section \ref{sec3} after presenting the proposed algorithm for CKM construction.

\section{Kriging Algorithm for CKM Construction}
\label{sec2}
This section focuses on the construction of CKMs, which is a challenging task when there is only a limited number of measurement points available. Without loss of generality, we consider one particular GBS $k$, and for notational simplicity, the index $k$ is suppressed from the CKM $\mathcal G_k$ and the measurement points $\{(\mathbf{x}_{i}^{(k)}, \tilde Z_k(\mathbf{x}_{i}^{(k)})), i=1, \dots ,N_k\}$. In particular, we propose a data-driven method based on the Kriging interpolation algorithm to efficiently construct CKMs for UAV communications in a 3D environment at a fixed altitude. In the following, Section \ref{sec2.2} first provides an overview of the Kriging interpolation method. Then, Section \ref{sec2.1} presents the semivariogram model that is used to learn the distribution of the measurement samples. Finally, Section \ref{sec2.3} uses the Kriging interpolation method to estimate the CKM based on the acquired semivariogram model and the measurement samples.
\subsection{Kriging Interpolation Algorithm}
\label{sec2.2}
Kriging is often adopted to find the best linear unbiased estimation of the channel knowledge among the measurement points \cite{N. Cressie}. In particular, Kriging can be regarded as a linear interpolation in a stochastic perspective, with the simplified model as
\begin{equation}
\label{kriging}
Z({\mathbf{x}})=\sum_{i=1}^{N} w_{i} \tilde Z(\mathbf{x}_{i}),
\end{equation}
where $Z({\mathbf{x}})$ is the estimated value of location $\mathbf{x} \in \mathcal{D}, \{\tilde Z(\mathbf{x}_{i}), i \in 1, \dots, N\}$ are the values of $N$ neighboring samples, and the weights $w_i$ are parameters to be designed to make the estimation in \eqref{kriging} unbiased with the minimum variance of estimation error.
\\\indent First, in order to make \eqref{kriging} an unbiased estimator, we need to ensure that the mean of the estimations must be equal to that of the ground-truth values, i.e.,
\begin{equation}
    \mathbb{E}[Z({\mathbf{x}})-\tilde{Z}({\mathbf{x}})]=\mathbb{E}[Z({\mathbf{x}})]-\mathbb{E}[{\tilde Z}({\mathbf{x}})]=0,
\end{equation}
where $\mathbb{E}[\cdot]$ denotes the mathematical expectation.
Towards this end, let $\epsilon_{\mathbf{x}}$ denote the estimation error for estimating ${Z}({\mathbf{x}})$, i.e.,
\begin{equation}
\epsilon_{\mathbf{x}}=Z({\mathbf{x}})-\tilde{Z}({\mathbf{x}})=\sum_{i=1}^{N} w_{i} \tilde Z(\mathbf{x}_{i})-\tilde{Z}({\mathbf{x}}).
\end{equation}
Then, the above two criteria can be expressed in terms of the mean and variance of $\epsilon_{\mathbf{x}}$. We consider that the channel knowledge has constant/stationary mean with $\mathbb{E}[\tilde Z(\mathbf{x}_{i})]=\mathbb{E}[\tilde{Z}({\mathbf{x}})]=m$, then we have
\begin{equation}
\begin{aligned}
\mathbb{E}[\epsilon_{\mathbf{x}}]=0 & \Leftrightarrow \sum_{i=1}^{N} w_{i} \mathbb{E}[\tilde Z(\mathbf{x}_{i})]-\mathbb{E}[\tilde{Z}({\mathbf{x}})]=0 \\
& \Leftrightarrow m \sum_{i=1}^{N} w_{i}-m=0 \\
& \Leftrightarrow \sum_{i=1}^{N} w_{i}=1,
\end{aligned}
\end{equation}
which indicates that the summation of the weights should be unity for the estimation to be unbiased.
\\\indent Next, we need to find the estimator where the estimation error has the minimum variance, since when the degree of dispersion between the estimated value and the ground-truth value is larger, the estimation is less accurate. Towards this end, we first derive the estimation variance $\operatorname{Var}(\epsilon_{\mathbf{x}})$, i.e.,
\begin{equation}
\begin{aligned}
 \operatorname{Var}(\epsilon_{\mathbf{x}})&=\operatorname{Var}\left[\sum_{i=1}^{N} w_{i} \tilde Z(\mathbf{x}_{i})-\tilde{Z}({\mathbf{x}})\right] \\
&= \operatorname{Var}\left[\sum_{i=1}^{N} w_{i} \tilde Z(\mathbf{x}_{i})\right]-2 \operatorname{Cov}\left[\sum_{i=1}^{N} w_{i} \tilde Z(\mathbf{x}_{i}), \tilde{Z}({\mathbf{x}})\right]+\operatorname{Var}(\tilde{Z}({\mathbf{x}})) \\
&= \sum_{i=1}^{N} \sum_{j=1}^{N} w_{i} w_{j} \operatorname{Cov}(\tilde Z(\mathbf{x}_{i}), \tilde Z(\mathbf{x}_{j}))-2 \sum_{i=1}^{N} w_{i} \operatorname{Cov}(\tilde Z(\mathbf{x}_{i}), \tilde{Z}({\mathbf{x}}))+\operatorname{Cov}(\tilde{Z}({\mathbf{x}}), \tilde{Z}({\mathbf{x}})),
\label{equ:var}
\end{aligned}
\end{equation}
where $\operatorname{Var}(\epsilon_{\mathbf{x}})$ denotes the variance of $\epsilon_{\mathbf{x}}$, and $\operatorname{Cov}(\tilde Z(\mathbf{x}_{i}), \tilde{Z}({\mathbf{x}}))$ denotes the covariance between $\tilde Z(\mathbf{x}_{i})$ and $\tilde{Z}({\mathbf{x}})$.
Then, we minimize the estimation variance $\operatorname{Var}(\epsilon_{\mathbf{x}})$ by solving the following optimization problem:
\begin{align}
        \text{(P3.1)}:&~\min\limits_{\{w_{i}\}}~\operatorname{Var}(\epsilon_{\mathbf{x}})\notag\\
        &~~~\text {s.t.} ~ \sum_{i=1}^{N} w_{i}=1.\notag
        \notag
\end{align}
To acquire the optimal weights of ${w_i}$ to problem (P3.1), we need to construct a semivariogram model as follows.
\subsection{Semivariogram Model}
\label{sec2.1}
The semivariogram is defined as a function to depict the spatial autocorrelation of sampled measurement points \cite{N. Cressie}.
First, the covariance of the channel knowledge $\tilde Z(\mathbf{x})$ at different locations is given as
\begin{equation}
    \begin{aligned}
\operatorname{Cov}(\mathbf{u}, \mathbf{v}) & \triangleq \operatorname{Cov}(\tilde Z(\mathbf{u}), \tilde Z(\mathbf{v})) \\
&=\mathbb{E}\{(\tilde Z(\mathbf{u})-\mathbb{E}[\tilde Z(\mathbf{u})])(\tilde Z(\mathbf{v})-\mathbb{E}[\tilde Z(\mathbf{v})])\}\\
&=\mathbb{E}[(\tilde Z(\mathbf{u}) \tilde Z(\mathbf{v}))]-\mathbb{E}[\tilde Z(\mathbf{u})] \mathbb{E}[\tilde Z(\mathbf{v})],
\end{aligned}
\end{equation}
where $\mathbf{u}, \mathbf{v} \in \mathcal D$.  Then, the semivariogram is defined as
\begin{equation}
\label{semi2.2}
    \begin{aligned}
\gamma(\mathbf{u}, \mathbf{v}) \triangleq & \frac{1}{2} \operatorname{Var}(\tilde Z(\mathbf{u})-\tilde Z(\mathbf{v})) \\
=& \frac{1}{2} \mathbb{E}\left\{((\tilde Z(\mathbf{u})-\mathbb{E}[\tilde Z(\mathbf{u})])-(\tilde Z(\mathbf{v})-\mathbb{E}[\tilde Z(\mathbf{v})]))^{2}\right\} \\
=& \frac{1}{2} \mathbb{E}\left\{(\tilde Z(\mathbf{u})-\mathbb{E}[\tilde Z(\mathbf{u})])^{2}\right\}+\frac{1}{2} \mathbb{E}\left\{(\tilde Z(\mathbf{v})-\mathbb{E}[\tilde Z(\mathbf{v})])^{2}\right\} \\
&-\mathbb{E}\{(\tilde Z(\mathbf{u})-\mathbb{E}[\tilde Z(\mathbf{u})])(\tilde Z(\mathbf{v})-\mathbb{E}[\tilde Z(\mathbf{v})])\},
\end{aligned}
\end{equation}
where $\text{Var}(\tilde Z(\mathbf{u})-\tilde Z(\mathbf{v}))$ denotes the variance of $\tilde Z(\mathbf{u})-\tilde Z(\mathbf{v})$. The semivariogram in \eqref{semi2.2} can be expressed in terms of the covariance function as
\begin{equation}
\label{equ:corre}
    \gamma(\mathbf{u}, \mathbf{v})=\frac{1}{2} \operatorname{Cov}(\mathbf{u}, \mathbf{u})+\frac{1}{2} \operatorname{Cov}(\mathbf{v}, \mathbf{v})-\operatorname{Cov}(\mathbf{u}, \mathbf{v}).
\end{equation}
Because $\tilde Z(\mathbf{x})$ follows a second-order stationary process, as a result, the covariance $\operatorname{Cov}(\mathbf{u},\mathbf{v})$ and semivariogram $\gamma(\mathbf{u},\mathbf{v})$ are both functions of $\mathbf{h} \triangleq \mathbf{v}-\mathbf{u}$. Then, the relationship in \eqref{equ:corre} can be simplified as
\begin{equation}
    \gamma(\mathbf{h})=C(\mathbf{0})-C(\mathbf{h}),
\end{equation}
where $C(\mathbf{0})$ is the autocovariance at lag $\mathbf{0}$, which is also known as the variance of the process, and $C(\mathbf{h})$ denotes the autocovariance at lag $\mathbf{h}$.
\\\indent Since that $\tilde Z(\mathbf{x})$ is isotropic, $\tilde Z(\mathbf{x})$ has same properties of covariance in all directions. $\gamma(\mathbf{h})$ and $C(\mathbf{h})$ become functions of the distance $h \triangleq\|\mathbf{h}\|$ only, which are denoted as $\gamma(h)$ and $C(h)$. In this case, the spatial analog of autocorrelation, known as {\it{correlogram}}, is defined as $\rho(h) \triangleq C(h) / C(0)$, which is referred to the correlation function. Furthermore, the semivariogram is re-expressed in a more descriptive way as
\begin{equation}
\label{semi2.5}
    \gamma(h)=\tau^{2}+\sigma^{2}(1-\rho_{0}(h / \omega)), \text { with } \tau^{2} \in \mathbb{R}_{0}^{+}, \text{ and } \sigma^{2}, \omega \in \mathbb{R}^{+},
\end{equation}
where $\mathbb{R}_{0}^{+}$ denotes non-negative real numbers, $\mathbb{R}^{+}$ denotes positive real numbers, and $\tau^{2}$ is the sampling error, which is also called {\it{nugget variance}} and usually should be zero. The variance of the process $C(0)=\sigma^{2}$ is also called {\it{sill}} since $\lim _{h \rightarrow \infty} \gamma(h)=\sigma^{2}$. The parameter $\omega$ defined as an additional scaling factor of the correlogram $\rho(h)=\rho_{0}(h / \omega)$ determines the {\it{range}} of the model.
\begin{figure}[t]
        \centering
        \includegraphics[scale=0.9]{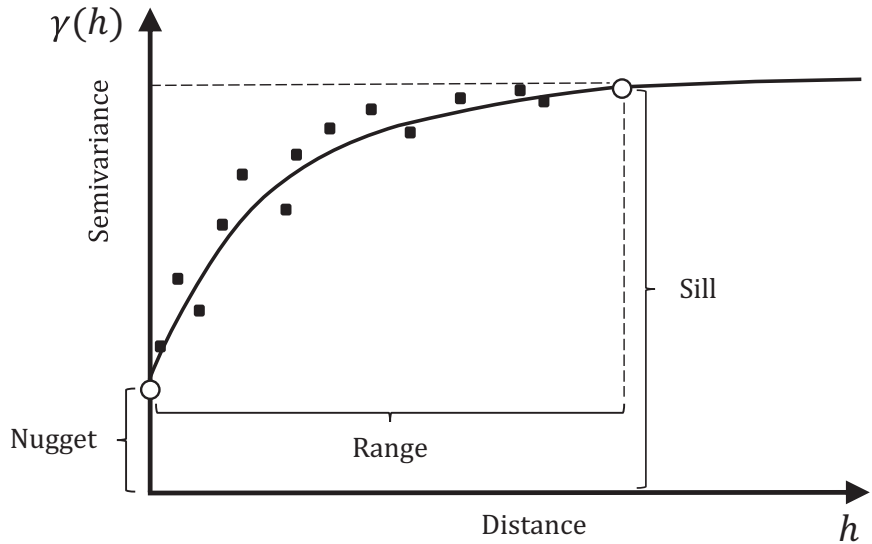}
        \caption{Illustration of parameters for describing semivariogram \cite{N. Cressie}.}
        \label{semivario}
\end{figure}
Fig. \ref{semivario} describes the fitting parameters that characterize the semivariogram, where the black dots represent sampling observations. The nugget parameter $\tau^{2}$ is the jump height of the semivariogram at the discontinuity at $h=0$. The sill parameter $\sigma^{2}$ characterizes the limit of the semivariogram when $h$ tends to infinity. Range $\omega$ indicates the distance in which the difference of the semivariogram is negligible as $h$ further increases, i.e., $\rho_{0}(h / \omega)=0$.
\\\indent Next, we use the semivariogram in \eqref{semi2.5} to represent the second-order statistics, which is used for the interpolation process in Section \ref{sec2.2}. Now, it remains for us to determine \eqref{semi2.5} based on the measurement points. Towards this end, we need to choose a proper mathematical model to be fitted in \eqref{semi2.5}.
Two common options are the {\it{exponential model}}
\begin{equation}
\label{expmodel}
\gamma_{\exp }(h)=a+b(1-\exp (-h / c)),
\end{equation}
and the {\it{spherical model}}
\begin{equation}
\gamma_{\text {spherical }}(h)=a+b(3 h / 2 c-\frac{1}{2}(h / c)^{3}),
\end{equation}
where $a,b,c \in \mathbb{R}^+$ are the modelling parameters. The parameters of the above models can be usually obtained by least-squares fitting or maximum likelihood estimation.
\subsection{Solution to Problem (P3.1)}
\label{sec2.3}
Based on the fitted semivariogram model, we can acquire the semivariance $\gamma(h)$ between any two given measurement data points, which can be used to substitute the unknown covariance between these two points with the semivariance value in problem (P3.1). To be specific, we define $\gamma_{ij} \stackrel{\Delta}{=} \sigma^{2}-C_{i j}$ with $C_{i j}=\operatorname{Cov}(\tilde Z(\mathbf{x}_{i}), \tilde Z(\mathbf{x}_{j}))$, and $\gamma_{i 0} \triangleq \sigma^{2}-C_{i 0}$ with $C_{i 0}=\operatorname{Cov}(\tilde Z(\mathbf{x}_{i}), \tilde Z({\mathbf{x}}))$. Then by substituting $\operatorname{Cov}(\tilde Z(\mathbf{x}_{i}), \tilde Z(\mathbf{x}_{j}))$ as $\sigma^{2}-\gamma_{i j}$ and $\operatorname{Cov}(\tilde Z(\mathbf{x}_{i}), \tilde Z({\mathbf{x}}))$ as $\sigma^{2}-\gamma_{i 0}$, the estimation variance in \eqref{equ:var} is rewritten as
\begin{equation}
\begin{aligned}
&\operatorname{Var}(\epsilon_{\mathbf{x}})=
&\sum_{i=1}^{N} \sum_{j=1}^{N} w_{i} w_{j}(\sigma^{2}-\gamma_{i j})-2 \sum_{i=1}^{N} w_{i}(\sigma^{2}-\gamma_{i 0})+(\sigma^{2}-\gamma_{00}).
\end{aligned}
\end{equation}
With $\sum_{i=1}^{N} w_{i}=1$, then we have
\begin{equation}
\begin{aligned}
\operatorname{Var}(\epsilon_{\mathbf{x}})
&=\sum_{i=1}^{N} \sum_{j=1}^{N} w_{i} w_{j}(\sigma^{2}-\gamma_{i j})-2 \sum_{i=1}^{N} w_{i}(\sigma^{2}-\gamma_{i 0})+(\sigma^{2}-\gamma_{00}) \\
&=2 \sum_{i=1}^{N} w_{i} \gamma_{i 0}-\sum_{i=1}^{N} \sum_{j=1}^{N} w_{i} w_{j} \gamma_{i j}-\gamma_{00} \\
&\triangleq J(\{w_{i}\}).
\end{aligned}
\end{equation}
Our objective is to minimize the estimation variance $\operatorname{Var}(\epsilon_{\mathbf{x}})$, which can be formulated as
\begin{align}
        \text{(P3.2)}:&~\min\limits_{\{w_{i}\}}~J(\{w_{i}\})\notag\\
        &~~~\text {s.t.} ~ \sum_{i=1}^{N} w_{i}=1.\notag
        \notag
\end{align}
By introducing the Lagrange multiplier $\mu$ for the constraint in problem (P3.2), the Lagrangian of problem (P3.2) is
\begin{equation}
\mathcal{L}(\{w_{i}\}, \mu)=J(\{w_{i}\})+2\mu(\sum_{i=1}^{N} w_{i}-1).
\end{equation}
Then, we can obtain the optimal weights by examining the Karush-Kuhn-Tucker (KKT) conditions with $\frac{\partial \mathcal{L}(\{w_{i}\}, \mu)}{\partial w_{i}}=0$, $i = 1, \dots, N$, and $\frac{\partial \mathcal{L}(\{w_{i}\}, \mu)}{\partial \mu}=0$. As a result, we have the following system of linear equations:
\begin{equation}
\label{equ:equsystem}
\left\{\begin{array}{l}
\sum_{j=1}^{N} w_{j} \gamma_{i j}-\mu-\gamma_{i 0}=0, \forall i \in\{1, \ldots, N\} \\
\sum_{i=1}^{N} w_{i}=1.
\end{array}\right.
\end{equation}
Note that with given fitted semivariogram ${\gamma_{ij}}$, the optimal $w_i$ is easy to be obtained via simultaneously solving the above $N+1$ linear equations in \eqref{equ:equsystem}. Suppose that the obtained solution to (P3.2) is denoted by $w^{*}_{i}, i=1, \dots, N$. As a result, the spatial interpolation model for constructing the CKM is obtained as
\begin{equation}
Z(\mathbf{x})=\sum_{i=1}^{N} w^{*}_{i} \tilde Z(\mathbf{x}_{i}).
\end{equation}
\section{Derivative-Free Placement Optimization Based on CKMs}
\label{sec3}
This section studies the UAV placement optimization problem in a multi-UAV wireless network by utilizing CKMs constructed in Section \ref{sec2}. Notice that since general CKMs normally contain discrete location-specific channel knowledge without closed-form expressions, the corresponding weighted sum rate function in (P1) becomes non-differentiable with respect to UAV locations. In this case, conventional optimization techniques relying on function derivatives no longer work. To address this issue, Section \ref{sec3.2} first provides a brief overview about the derivative-free optimization. Then, Section \ref{UAV:Opti} presents a novel iterative algorithm based on the derivative-free optimization framework to solve the non-differentiable weighted sum rate maximization problem (P1) due to the discrete channel knowledge from the constructed CKMs.

\subsection{Derivative-free optimization}
\label{sec3.2}
We first provide a brief review about derivative-free optimization \cite{marazzi2002wedge}. The purpose of derivative-free optimization is to maximize an objective function $f(\mathbf{x})$ with variable $\mathbf{x} \in \mathbb{R}^{n\times1}$, where the derivative of $f(\mathbf{x})$ is not available. The basic idea is to iteratively approximate $f(\mathbf{x})$ by a series of analytic functions under interpolation conditions, and accordingly update the optimization variable by maximizing the approximate function in each iteration, subject to a trust region constraint.
\\\indent In particular, consider one iteration $i \ge 1$, in which the local point is denoted by $\mathbf{x}^{(i)}$. First, we construct an analytic function $\phi^{(i)}(\mathbf{x})$ (with $m$ parameters to be determined) to approximate the objective function $f(\mathbf{x})$ near the local point $\mathbf{x}^{(i)}$. In practice, the quadratic and linear functions are widely adopted for constructing $\phi^{(i)}(\mathbf{x})$ \cite{marazzi2002wedge}, which include $m=\frac{1}{2}(n+1)(n+2)$ and $m=n+1$ parameters, respectively. To determine these function parameters, we introduce a so-called \textit{interpolation set} $\bm{\Sigma}$ containing $m-1$ points, and accordingly impose the corresponding interpolation conditions, i.e.,
\begin{align}
\phi^{(i)}(\mathbf{y})=f(\mathbf{y}), \forall \mathbf{y} \in \{\mathbf{x}^{(i)}\} \cup \bm{\Sigma}.\label{intercond}
\end{align}
Notice that points in the interpolation set are randomly generated initially, and will be updated as the iteration proceeds. Also notice that the constructed interpolation set $\bm{\Sigma}$ should be {\it non-degenerate}, i.e., based on the equations in \eqref{intercond} the parameters in the approximate function $\phi^{(i)}(\mathbf{x})$ is non-singular, such that $\phi^{(i)}(\mathbf{x})$ can be uniquely determined.
\\\indent Next, we update the optimization variable. Towards this end, we first find a trial point as $\mathbf{x}^{(i)}_{+}=\mathbf{x}^{(i)}+\mathbf{s}^{(i)}$, where the update step $\mathbf{s}^{(i)}$ is obtained by maximizing the approximate function $\phi^{(i)}(\mathbf{x}^{(i)}+\mathbf{s})$ subject to a newly imposed trust region, i.e.,
\begin{align}
        \mathbf{s}^{(i)} = \arg&\max\limits_{\mathbf{s}}~\phi^{(i)}(\mathbf{x}^{(i)}+\mathbf{s})\notag\\
       &~~\text {s.t.} ~~~ \|\mathbf{s}\| \leq \Delta,\notag
\end{align}
where $\Delta$ is the trust region size that is set to a given value $\Delta_0$ at the beginning of the algorithm. If the resultant function value increases (i.e., $f(\mathbf{x}^{(i)}_+) >  f(\mathbf{x}^{(i)})$), then we update the variable as $\mathbf{x}^{(i+1)} =  \mathbf{x}^{(i)}_+$ for the next iteration; otherwise, we have $\mathbf{x}^{(i+1)} =  \mathbf{x}^{(i)}$. In either case, we update the interpolation set $\bm{\Sigma}$ by adding $\mathbf{x}^{(i)}_+$ as a new point and removing an old point that is furthest from $\mathbf{x}^{(i)}_+$, and also decrease the trust region size $\Delta$ to $\beta \Delta$ with $0< \beta <1$.

Notice that the above iteration terminates when the local point and interpolation set converges (i.e., $\| \mathbf{x}^{(i+1)} - \mathbf{y} \| \le \epsilon, \forall \mathbf{y} \in \bm{\Sigma}$) and $\Delta < \epsilon$ is satisfied at the same time, or the maximum number of iterations is met, where $\epsilon$ is a sufficiently small constant threshold for determining convergence. Nevertheless, if $\Delta < \epsilon $ is met but the local point and interpolation set do not converge yet, then we should reset $\Delta$ as $\Delta = \Delta_0$ and run the next iteration.
\subsection{Derivative-free UAV placement optimization}
\label{UAV:Opti}
Building upon the derivative-free optimization, in this section we develop an efficient derivative-free algorithm for solving problem (P1), which is implemented in an iterative manner. For notational convenience, we denote the objective function in problem (P1) as $f(\{\mathbf{q}_{j}\}) = \sum\nolimits_{k\in{\cal K}}\alpha_{k}r_{k}(\{\mathbf{q}_{j}\})$.

In particular, consider any given iteration $i \geq 1$, where the local point is given by $\{\mathbf{q}^{(i)}_j\}$. We denote the locations of the $K$ UAVs $\mathbf{q}^{(i)}_c=(\mathbf{q}^{(i)}_{1},\mathbf{q}^{(i)}_{2},...,\mathbf{q}^{(i)}_{K})^{T} \in \mathbb{R}^{2K\times1}$ for notational convenience. First, we approximate the objective function $f(\{\mathbf{q}_{j}\})$ by the function $\phi^{(i)}(\mathbf{q}^{(i)}_{c}+\mathbf{s})$. In particular, we adopt quadratic function for $\phi^{(i)}(\mathbf{q}^{(i)}_{c}+\mathbf{s})$, given by
\begin{equation}
    \phi^{(i)}(\mathbf{q}^{(i)}_{c}+\mathbf{s})=f(\mathbf{q}^{(i)}_{c})+{\mathbf{g}^{(i)}}^T \mathbf{s}+\frac{1}{2} {\mathbf{s}}^{T} \mathbf{G}^{(i)} \mathbf{s},\label{quarfunction}
\end{equation}
where the vector $\mathbf{g}^{(i)} \in \mathbb{R}^{2K\times1}$ and the symmetric matrix $\mathbf{G}^{(i)} \in \mathbb{R}^{2K\times2K}$ contain $m-1=\frac{1}{2}(2K+1)(2K+2)-1$ parameters to be determined. Notice that the consideration of quadratic function in \eqref{quarfunction} is due to the fact that it can properly balance between the approximation performance and the computation burden.
To uniquely determine $\phi^{(i)}(\mathbf{q}^{(i)}_c+\mathbf{s})$, we need to find a non-degenerate interpolation set of $m-1$ points, denoted by
\begin{equation}
    \bm{\Sigma}=\{\mathbf{y}_{1}, ..., \mathbf{y}_{m-1}\},
\end{equation}
\text{where} $\mathbf{y}_{l} \in \mathbb{R}^{2K\times1}, ~ l=1, ..., m-1$. Note that points in the interpolation set $\bm{\Sigma}$ are obtained through randomly sampling from a uniform distribution over region ${\mathcal D}$ in \eqref{range} and will be updated in each iteration.
Accordingly, we have the following $m$ equalities:
\begin{equation}
 \label{interpo}
    \phi^{(i)}(\mathbf{q}^{(i)}_{c})=f(\mathbf{q}^{(i)}_{c}), ~ \phi^{(i)}(\mathbf{y}_{l})=f(\mathbf{y}_{l}), ~ l=1, ..., m-1.
\end{equation}
By solving the system of linear equations, we obtain $\mathbf{g}^{(i)}$ and $\mathbf{G}^{(i)}$ and accordingly determine the approximate function $\phi^{(i)}(\mathbf{q}^{(i)}_c+\mathbf{s})$.

\indent Next, we update the UAVs' placement locations. Towards this end, we first obtain a trial step $\mathbf{s}^{(i)}$, by solving the following problem (P4.1).
\begin{align}
        \text{(P4.1)}:&~\max\limits_{\mathbf{s}}~\phi^{(i)}(\mathbf{q}^{(i)}_{c}+\mathbf{s})\notag\\
        &~~~\text {s.t.} ~ \|\mathbf{s}\| \leq \Delta\notag\\
        &~~~~~~~~\mathbf{q}^{(i)}_{c}+\mathbf{s} \in \mathcal{D}.\notag
\end{align}
Note that as the symmetric matrix $\mathbf{G}^{(i)}$ may not be negative semi-definite, problem (P4.1) may be a non-convex quadratic program in general that is difficult to be optimally solved. In order to solve problem (P4.1), we utilize a trust region subproblem solver tool in \cite{toolbox}, which applies the subroutine Trust Region Step in the BOX (TRSBOX) of BOBYQA algorithm \cite{BOBYQA}. As a result, we approximately obtain the trial step $\mathbf{s}^{(i)}$ and accordingly obtain the trial point $\mathbf{q}^{(i)}_+=\mathbf{q}^{(i)}_{c}+\mathbf{s}^{(i)}$.
\\\indent With the trial step $\mathbf{s}^{(i)}$ at hand, we are ready to update the UAV placement locations $\mathbf{q}_c^{(i+1)}$, together with the trust region size $\Delta$ and the interpolation set $\bm{\Sigma}$. Towards this end, we obtain $\mathbf{y}_{l_{\text {out }}}=\arg\max\nolimits_{\mathbf{y} \in \bm{\Sigma}}\|\mathbf{y}-\mathbf{q}_{c}^{(i)}\|$ as the point in $\bm{\Sigma}$ with the longest distance from the local point $\mathbf{q}^{(i)}_c$, which is a candidate point to be removed from $\bm{\Sigma}$.
In particular, if the trial point $\mathbf{q}^{(i)}_+$ leads to a weighted sum rate $f(\{\mathbf{q}_{k}\})$ that is greater than $\mathbf{q}^{(i)}_c$, i.e., $f(\mathbf{q}^{(i)}_{c}+\mathbf{s}^{(i)})>f(\mathbf{q}^{(i)}_{c})$, then we update the UAV placement locations as $\mathbf{q}^{(i+1)}_c=\mathbf{q}^{(i)}_c+\mathbf{s}^{(i)}$, which is also used as
the local point in the next iteration $i + 1$; otherwise, we have $\mathbf{q}^{(i+1)}_c=\mathbf{q}^{(i)}_c$.
Furthermore, we update the interpolation set $\bm{\Sigma}$ by adding the trial point $\mathbf{q}^{(i)}_+$ as a new point and removing $\mathbf{y}_{l_{\text {out}}}$, and also decrease $\Delta$ by factor $\beta$, i.e., $\Delta \gets \beta \Delta$.
The above iteration terminates when the local point and interpolation set converges (i.e., $\| \mathbf{q}_c^{(i+1)} - \mathbf{y} \| \le \epsilon, \forall \mathbf{y} \in \bm{\Sigma}$) and $\Delta < \epsilon$ is satisfied at the same time, or the maximum number of iterations is met. Nevertheless, if $\Delta < \epsilon$ is met but the local point and interpolation set do not converge yet, then we should reset the trust region size as $\Delta = \Delta_0$ and run the next iteration.
\\\indent In summary, we present the complete algorithm as Algorithm \ref{algorithm}. It is observed that Algorithm 1 results in monotonically non-decreasing objective function values, and the points in the interpolation set $\bm{\Sigma}$ will finally converge to the UAV placement locations. Therefore, the convergence of the algorithm can always be ensured.
It is worth comparing the complexity of the proposed derivative-free placement optimization design versus the optimal exhaustive search benchmark, in which we first sample the interested region ${\mathcal D}$ into $MN$ grids and then compare the weighted sum rates achieved by all the $MN$ possible UAV placement locations to get the desired solution. For the proposed algorithm, the total complexity of constructing the analytic function and solving problem (P4.1) for updating trial points is $\mathcal{O}(K^{4})$ \cite{BOBYQA}. For exhaustive search, the complexity is $\mathcal{O}(M^KN^{K})$. It is observed that when $K$ becomes large, the complexity of the proposed algorithm is much lower than that of the exhaustive search.
\begin{algorithm}[t]
    \caption{for Solving Problem (P1)}
    \label{algorithm}
    \begin{algorithmic}[1]
    \State Set the initial trust region size $\Delta_{0}>0$, and the initial UAV placement locations $\mathbf{q}^{(0)}_c$. Randomly generate $m-1$ points to compose an initial non-degenerate interpolation set $\bm{\Sigma}$. Set the iteration index as $i = 0$ and set the convergence threshold $\epsilon > 0$.
    \Repeat
    \State Construct the approximate quadratic function $\phi^{(i)}(\mathbf{q}^{(i)}_{c}+\mathbf{s})$ based on interpolation conditions in \eqref{interpo}.
    \State Compute the trial step $\mathbf{s}^{(i)}$ by solving problem (P3.2), and accordingly obtain the trial point $\mathbf{q}^{(i)}_+=\mathbf{q}^{(i)}_{c}+\mathbf{s}^{(i)}$.
    \State Find $\mathbf{y}_{l_{\text {out }}}=\arg\max\nolimits_{\mathbf{y} \in \bm{\Sigma}}\|\mathbf{y}-\mathbf{q}_{c}^{(i)}\|$.
    \If{$f(\mathbf{q}^{(i)}_{c}+\mathbf{s}^{(i)})>f(\mathbf{q}^{(i)}_{c})$}
    \State set $\bm{\Sigma}=\{\mathbf{q}^{(i)}_c\} \cup \bm{\Sigma} \backslash\{\mathbf{y}_{l_{\text {out }}}\}$, $\mathbf{q}^{(i+1)}_{c}=\mathbf{q}^{(i)}_{c}+\mathbf{s}^{(i)}$.
    \Else
    \State set $\Delta=\beta \Delta$, $\mathbf{q}^{(i+1)}_{c}=\mathbf{q}^{(i)}_{c}$.
    \If{$\|\mathbf{y}_{l_{\text {out }}}-\mathbf{q}^{(i)}_{c}\| \geq \|(\mathbf{q}^{(i)}_{c}+\mathbf{s}^{(i)})-\mathbf{q}^{(i)}_{c}\|$}
    \State set $\bm{\Sigma}=\{\mathbf{q}^{(i)}_{c}+\mathbf{s}^{(i)}\} \cup \bm{\Sigma} \backslash\{\mathbf{y}_{l_{\text {out }}}\}$.
    \EndIf \State {\textbf{end}}.
    \EndIf \State {\textbf{end}}.
    \If{$\Delta < \epsilon$, and $\| \mathbf{q}^{(i+1)}_c - \mathbf{y} \| >\epsilon, \forall \mathbf{y} \in \bm{\Sigma}$}
    \State reset $\Delta = \Delta_0$.
    \EndIf \State {\textbf{end}}.
    \State $i \leftarrow i+1.$
    \Until{$\Delta < \epsilon$, and $\| \mathbf{q}^{(i)}_c - \mathbf{y} \| \leq \epsilon, \forall \mathbf{y} \in \bm{\Sigma}$.}
    \end{algorithmic}
    \end{algorithm}
\section{Numerical Results}
\label{sec4}
This section presents numerical results to validate the performance of Kriging-based CKM constructions and the proposed derivative-free UAV placement optimization design with ground-truth CKMs and Kriging-estimated CKMs, as compared to some other benchmarks.
We consider a specific area in central Shanghai with a size of 300$\times$300 $\text{m}^2$, which consists of a dozen of buildings from a city map database.\footnote{The 3D city map is obtained online from \url{https://www.openstreetmap.org.}} The UAV altitude is set to be $H=50$ m. In the simulation, the received noise power at each GBS $k \in \cal K$ is $\sigma^{2}_k=-100~\mathrm{dBm}$, and the transmit power of each UAV $k \in \cal K$ is $P_k=30~\mathrm{dBm}$. The height of GBSs is set as $\hat{H}=2.0$ m, and the horizontal locations of GBSs are $\mathbf{w}_{1}=(-89.54~\text{m}, 16.30~\text{m})$, $\mathbf{w}_{2}=(-118.22~\text{m}, -53.86~\text{m})$, and $\mathbf{w}_{3}=(-18.15~\text{m}, -80.22~\text{m})$, respectively. The Remcom Wireless Insite software \cite{WI} is used to generate the CKM dataset at the GBS based on the 3D city map as the ground-truth, which provides true channel power gain values at uniformly distributed points. The CKM collects channel power gains at a total of 4347 points which are uniformly distributed every 5 m in both X- and Y-axis.
\subsection{Kriging-Based CKM Constructions}
\label{sec:construct}
We first present numerical results to show the performance of the Kriging algorithm for CKM constructions. Based on the ground-truth CKM, we uniformly sample each measurement point every 75 m along the X-axis at each coordinate in Y-axis and sample $N=300$ measurement points in total for data interpolation. For the Kriging algorithm, we choose the exponential model in \eqref{expmodel} as the semivariogram model.
\\\indent We consider the following two benchmark schemes for performance comparison:
\begin{itemize}
    \item {\it KNN} \cite{knn}: This scheme constructs the channel power gain $Z(\mathbf{x})$ at $\mathbf{x} \in \mathbb{R}^{1\times2}$ based on the same measurement samples as the Kriging algorithm. This scheme selects the 5 samples nearest to $\mathbf{x}$ and form the neighbor set $\mathcal{N}(\mathbf{x})$. Then, compute $Z(\mathbf{x})=\frac{1}{|\mathcal{N}(\mathbf{x})|}{\sum_{i \in |\mathcal{N}(\mathbf{x})|} Z(\mathbf{x}_i)}$.
    \item {\it Simplified LoS path loss channel model}: The LoS path channel model is considered in this scheme, for which the channel power gain between each point of interest $\mathbf{x} \in \mathbb{R}^{1\times2}$ and the GBS is given by $Z(\mathbf{x})={\beta_{0}}/{(\|\mathbf{x}-\mathbf{w}_{1}\|^{2}+(\hat{H}-H)^2)}$,
    where $\beta_{0}$ denotes the channel power gain at a reference distance of $d_0=1$ m and is set as $\beta_0=-30$ dB.
\end{itemize}
\begin{figure*}[!htb]
    \centering
    \subfigure[Tue CKM of GBS 1.]{
    \includegraphics[scale=0.585]{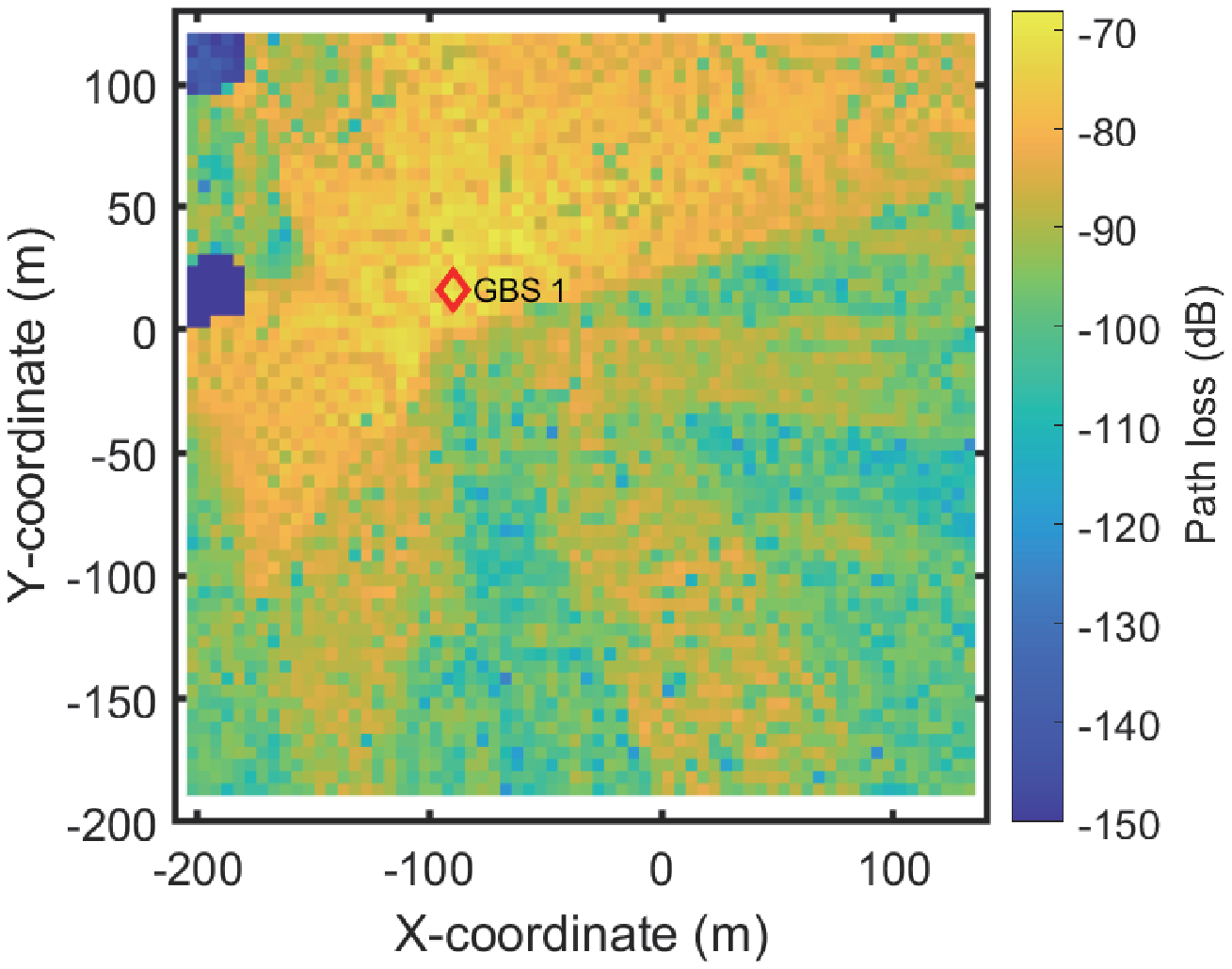}}
    \subfigure[Kriging-based CKM of GBS 1.]{
    \includegraphics[scale=0.585]{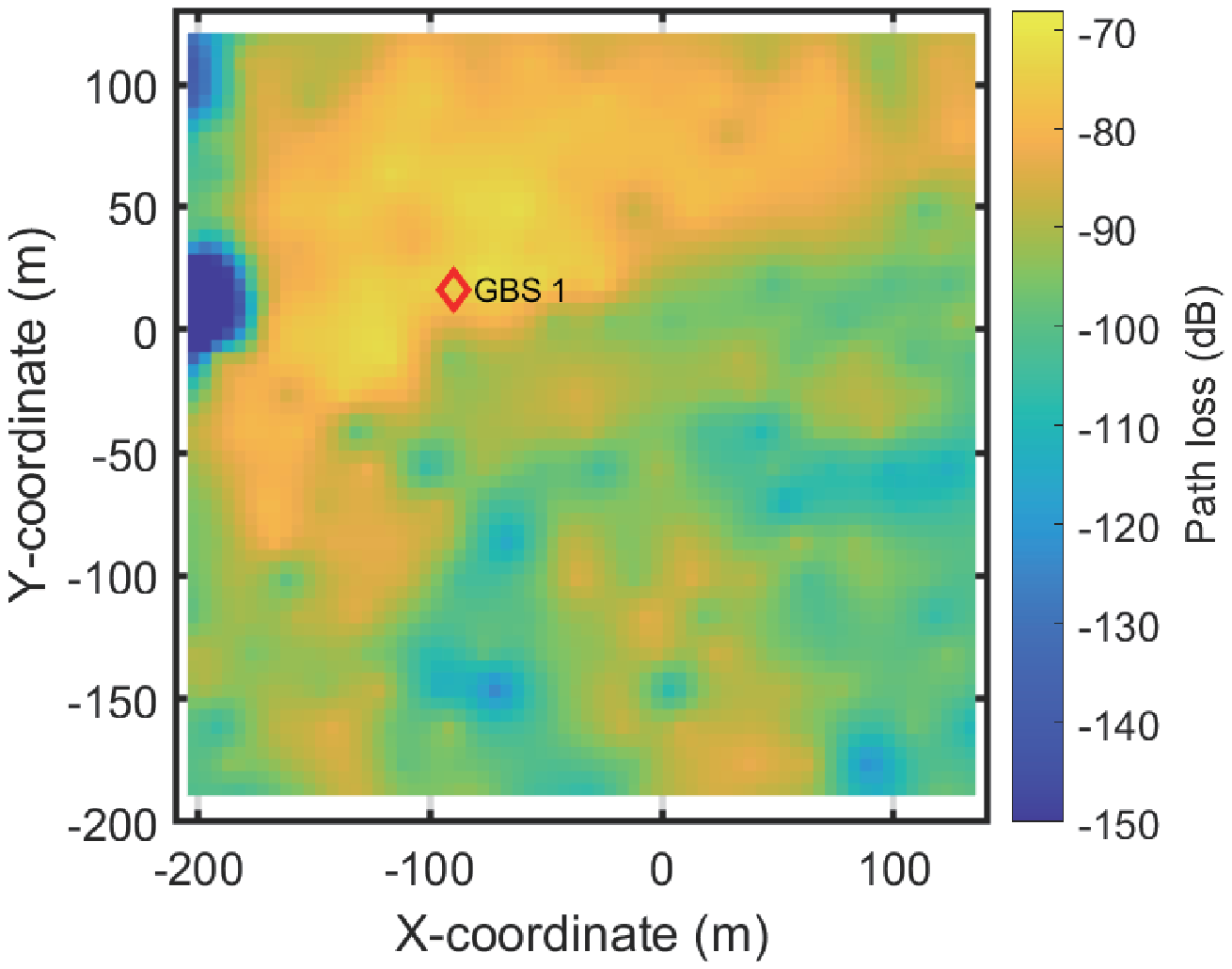}}
    \subfigure[KNN-based CKM of GBS 1.]{
    \includegraphics[scale=0.585]{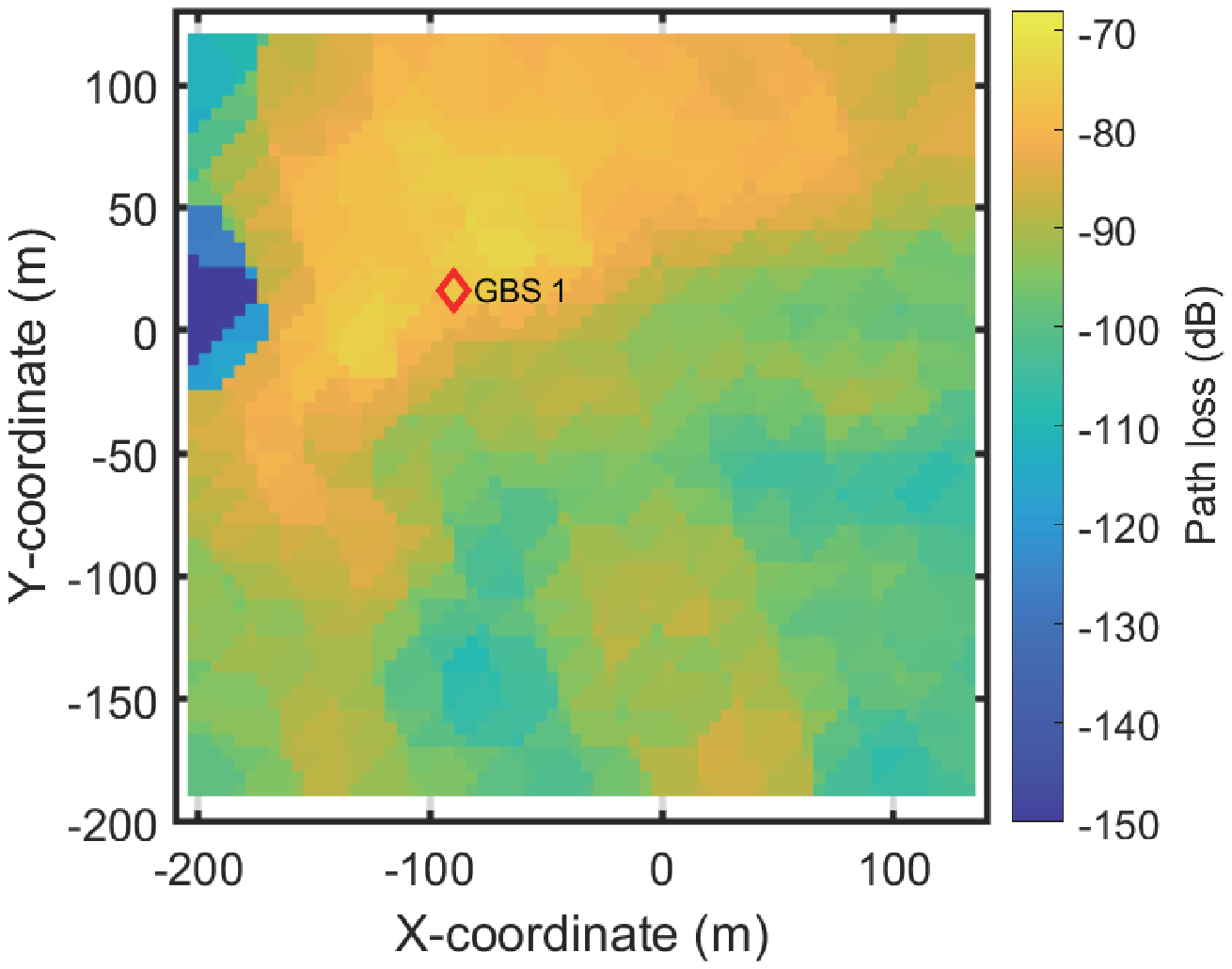}}
    \subfigure[Path loss model-based CKM of GBS 1.]{
    \includegraphics[scale=0.585]{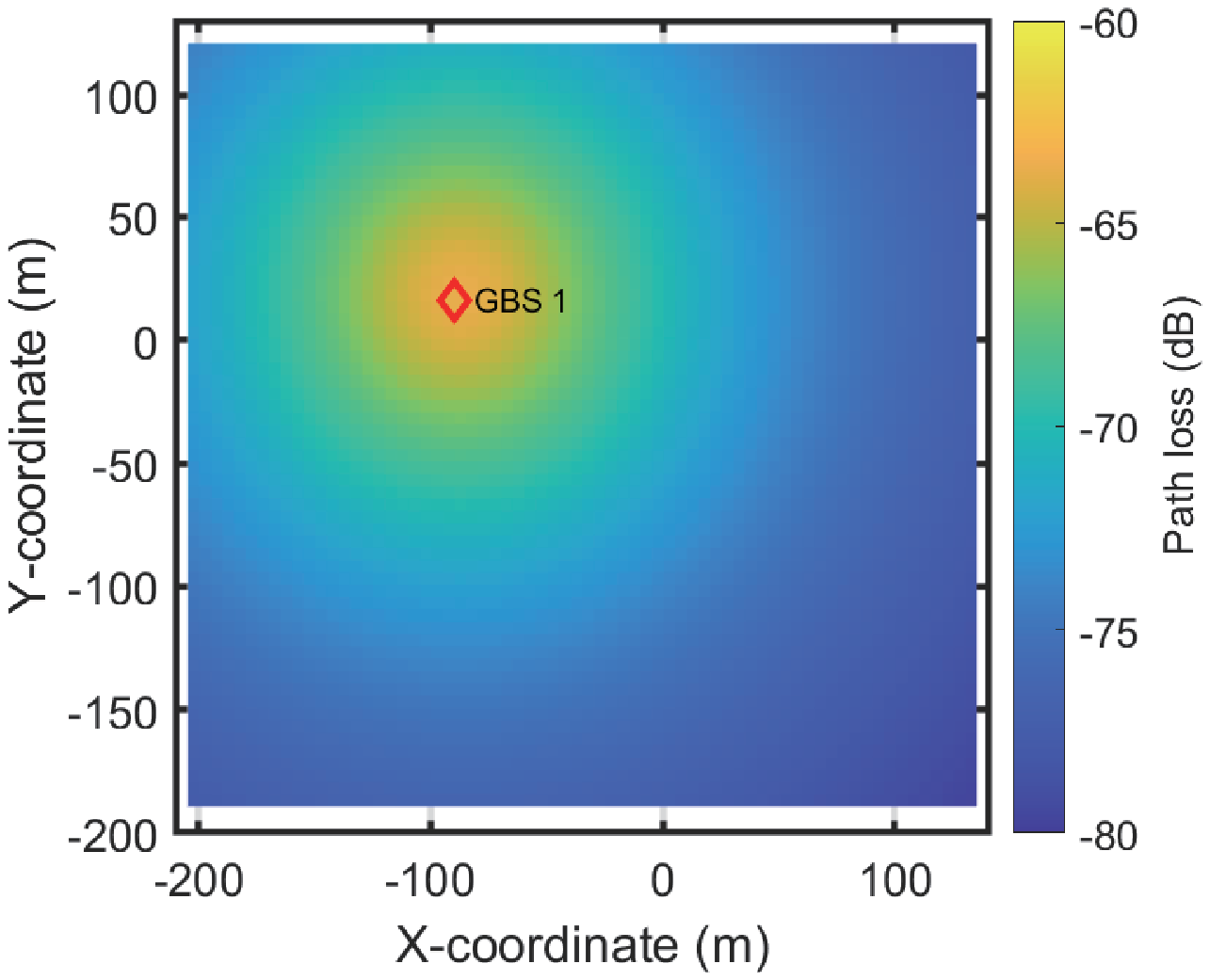}}
    \caption{CKM construction.}
    \label{fig:CKM-construct}
\end{figure*}
\indent \indent Fig. \ref{fig:CKM-construct} shows the ground-truth CKM and the construction results of the Kriging-based method together with the two aforementioned benchmark schemes. It is observed that with limited sample points for data interpolation, the Kriging-based method is able to estimate the full CKM with a distribution of channel power gains very close to the ground-truth. In addition, for the {\it{KNN}}-based method, it is observed that the constructed CKM is lack of details, and the areas of poor channel quality are not accurately estimated as compared with the Kriging-estimated CKM and the ground-truth CKM. In contrast, the {\it{simplified LoS path loss channel model}}-based CKM in Fig. \ref{fig:CKM-construct}(d) is observed as a concentric contour for channel power gains that only depend on the distance between the GBS and the points of interest, instead of their specific locations. This is quite different from the actual situation as shown in Fig. \ref{fig:CKM-construct}(a), as well as the data-driven CKM constructions in Figs. \ref{fig:CKM-construct}(b) and \ref{fig:CKM-construct}(c).
\\\indent Next, we consider the estimation performance of the Kriging-based method through calculating the corresponding mean absolute error (MAE) $e=\mathbb{E}[|\hat{Z}(\mathbf{x})-Z(\mathbf{x})|]$, where $\hat{Z}(\mathbf{x})$ and $Z(\mathbf{x})$ denote the estimated and ground-truth channel power gain value at point $\mathbf{x} \in \mathbb{R}^{1\times2}$.
\begin{figure}[h]
        \centering
        \includegraphics[scale=0.9]{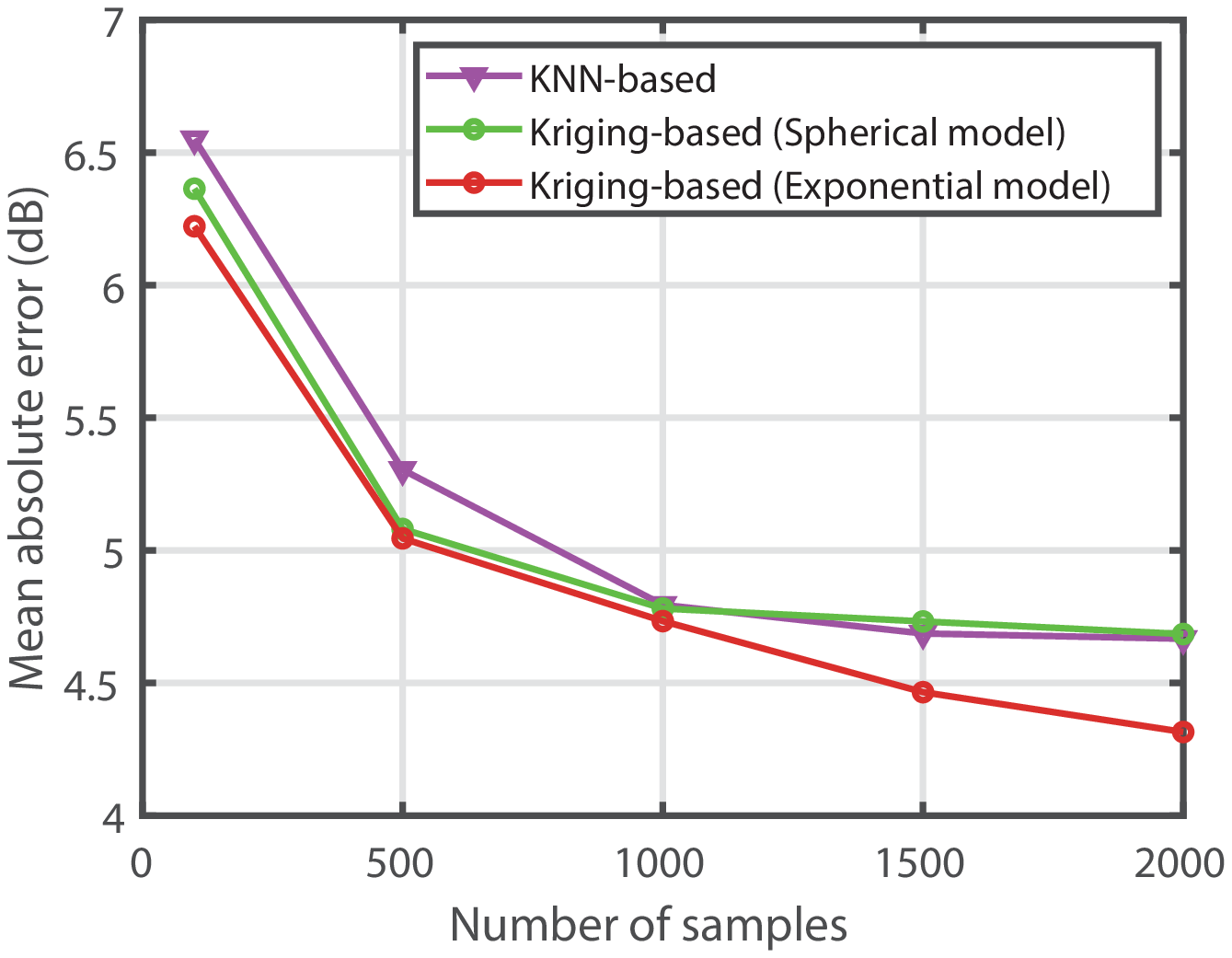}
        \caption{Construction error versus the number of samples.}
        \label{construct_error}
\end{figure}
Fig. \ref{construct_error} shows the construction error of the Kriging-based method with different semivariogram models and {\it{KNN}}-based method versus different numbers of samples. It is observed that the construction error of Kriging-based and {\it{KNN}}-based methods decreases as the number of samples increases, since more channel knowledge is available for more accurate CKM constructions. It is also observed that the Kriging-based method with the exponential model achieves the smallest construction errors among the three methods, and it only requires $1000$ samples to achieve a similar performance to that achieved by {\it{KNN}}-based method based on $2000$ samples. Furthermore, the construction error of the Kriging-based method with the exponential model still decreases while that of {\it{KNN}}-based method stables as the number of samples exceeds $1000$.
\subsection{Derivative-Free UAV Placement Optimization Design with CKMs}
We then present numerical results to show the performance of the derivative-free placement optimization design by considering both the ground-truth CKMs and the CKMs constructed based on the Kriging algorithm with the exponential model. The rate weights in problem (P1) are set as $\alpha_k=1,\forall k\in{\cal K}$, and thus the sum rate of UAVs is considered as the performance metric. For the proposed algorithm, the decreasing factor of the trust region is adopted as $\beta=0.5$ \cite{marazzi2002wedge}. For the CKM construction method based on the Kriging algorithm, we uniformly sample each measurement point from the ground-truth CKM for each GBS every 50 m along the X-axis at each coordinate in Y-axis. We collect $N=450$ measurement points in total for each GBS.
\\\indent We consider the following benchmark schemes with different placement design approaches and CKM construction methods for performance comparison.
\begin{itemize}
    \item {\it Exhaustive search}: For exhaustive search, the candidate UAV locations are uniformly sampled every 5 m, which is consistent with that in the CKM. This scheme calculates the weighted sum rate based on the ground-truth CKMs, and compares the resultant values to find the optimal placement locations.
    \item {\it Hovering above GBSs}: Each UAV hovers exactly above its associated GBS with $\mathbf{q}_k = \mathbf{w}_k, \forall k \in \cal K$. This scheme is generally optimal for the special case with $K=1$ or the inter-UAV interference is negligible.
    \item {\it Conventional design with LoS channels}: The LoS path channel model is considered in this scheme, for which the channel power gain between each UAV $j \in \cal K$ and GBS $k \in \cal K$ is given by $ Z_{k, j}(\mathbf{q}_{j})={\beta_{0}}/{(\|\mathbf{q}_{j}-\mathbf{w}_{k}\|^{2}+(\hat{H}-H)^2)}$,
    where $\beta_{0}$ denotes the channel gain at a reference distance of $d_0=1$ m and is set as $\beta_0=-30$ dB. This scheme corresponds to solving problem (P1) via conventional non-convex optimization techniques, such as successive convex approximation (see, e.g., \cite{UAV0, QQ2}).
\end{itemize}
Besides the ground-truth CKMs and the Kriging-estimated CKMs, we also consider the KNN-estimated CKMs for performance comparison, as introduced in Section \ref{sec:construct}.
\begin{figure}[ht]
    \centering
    \includegraphics[scale=0.9]{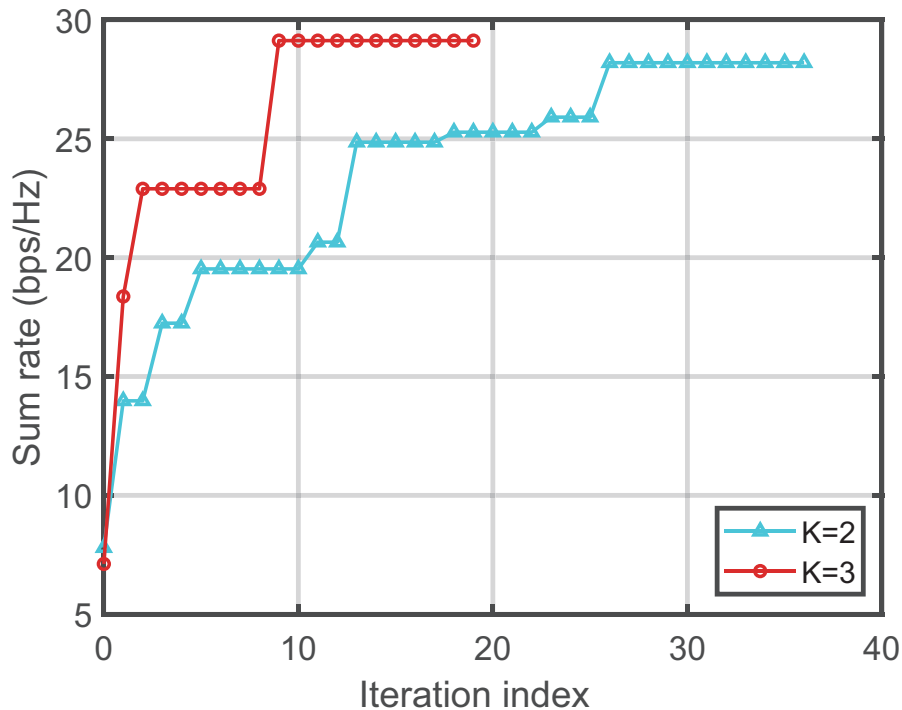}
    \caption{Convergence of the proposed algorithm.}
    \label{convergence}
\end{figure}
\\\indent Fig. \ref{convergence} shows the convergence of the proposed algorithm based on ground-truth CKMs, in both cases with $K = 2$ and $K = 3$. It is observed that the proposed algorithm takes 36 iterations and 19 iterations to converge for the case with $K=2$ and $K=3$, respectively. Furthermore, the proposed algorithm takes about $330$ and $930$ milliseconds to converge for the case with $K=2$ and $K=3$, respectively. For comparison, the {\it{exhaustive search}} takes about $60$ seconds and 166 hours to find the optimal UAV placement locations for the case with $K=2$ and $K=3$, respectively.
\begin{figure*}[!htb]
    \centering
    \subfigure[CKM of GBS 1.]{
    \includegraphics[scale=0.75]{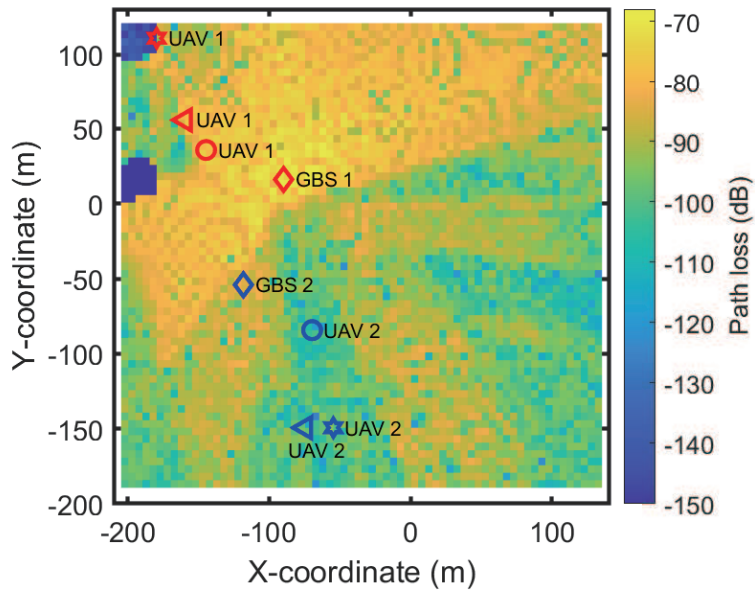}}
    \subfigure[CKM of GBS 2.]{
    \includegraphics[scale=0.75]{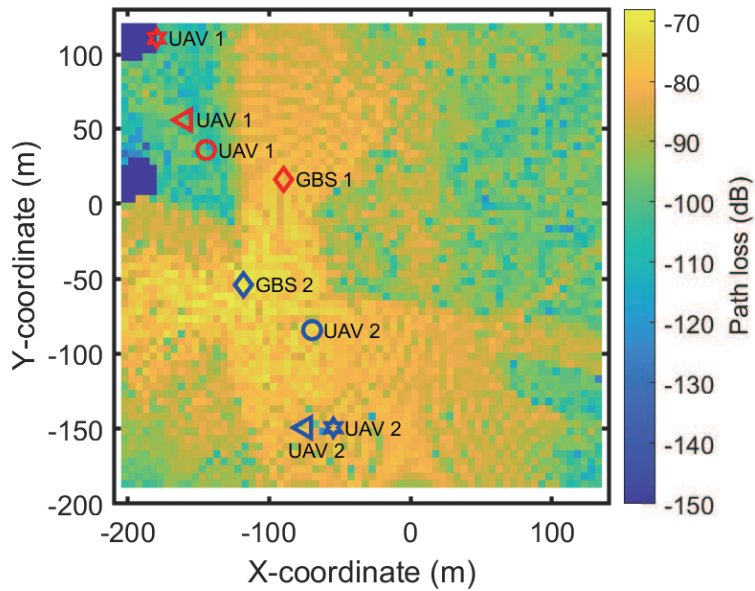}}
    \caption{Optimized UAV horizontal locations for the case with $K=2$. }
    \label{fig:CGMs}
\end{figure*}
\\\indent Fig. \ref{fig:CGMs} shows the optimized locations achieved by the proposed algorithm based on ground-truth, Kriging-estimated, and {\it{KNN}}-estimated CKMs for the case with $K=2$, where the diamonds indicate GBS locations, the circles, triangles, and hexagons indicate the UAV locations based on ground-truth, Kriging-estimated, and {\it{KNN}}-estimated CKMs, respectively. For the proposed algorithm based on ground-truth CKMs, it is observed that the horizontal location of UAV 1 converges to $\mathbf{q}^{*}_{1}=(-144.64~\text{m}, 35.69~\text{m})$, and that of UAV 2 converges to $\mathbf{q}^{*}_{2}=(-69.64~\text{m}, -84.31~\text{m})$. The sum rate converges to $R=28.201~\mathrm{bps/Hz}$, with the individual rates $r_1 = 13.572~\mathrm{bps/Hz}$ and $r_2 = 14.629~\mathrm{bps/Hz}$ for the two UAV users, respectively. In addition, for the {\it{hovering above GBSs}} scheme, the resulting sum rate is $R=7.799~\mathrm{bps/Hz}$, with the individual rates $r_1 = 3.473~\mathrm{bps/Hz}$, and $r_2 = 4.326~\mathrm{bps/Hz}$. It is observed that if the UAV users hover above their respectively associated GBSs, then both UAVs suffer severe co-channel interference from each other. After placement optimization, each UAV is observed to be placed at a location where its desired link between the corresponding GBS enjoys good channel quality, while its interference link between the other GBS is weak, thus mitigating the co-channel interference to the other UAV thus enhancing the SINR. Furthermore, for {\it{exhaustive search}}, the resulting sum rate is $R=28.201~\mathrm{bps/Hz}$, which are exactly the same with our proposed algorithm, and the UAV placement locations are also exactly the same.
\\\indent It is also interesting to compare the performance of the proposed algorithm based on ground-truth CKMs with that based on Kriging-estimated and {\it{KNN}}-estimated CKMs, as shown in Fig. \ref{fig:CGMs}. It is observed that for the proposed algorithm based on Kriging-estimated CKMs, the horizontal location of UAV 1 converges to $\mathbf{q}^{Kriging}_{1}=(-159.64~\text{m}, 55.69~\text{m})$ which deviates from the optimal location of UAV 1 based on the ground-truth CKMs with a distance of 25 m, and that of UAV 2 converges to $\mathbf{q}^{Kriging}_{2}=(-74.64~\text{m}, -149.31~\text{m})$ which is about 65.2 m away from the optimal location of UAV 2 based on the ground-truth CKMs. The sum rate converges to $R^{Kriging}=23.794~\mathrm{bps/Hz}$, with the individual rates $r^{Kriging}_1 = 14.027~\mathrm{bps/Hz}$ and $r^{Kriging}_2 = 9.767~\mathrm{bps/Hz}$ for the two UAV users, respectively. In addition, it is observed that for the proposed algorithm based on {\it{KNN}}-estimated CKMs, the horizontal location of UAV 1 converges to $\mathbf{q}^{KNN}_{1}=(-179.64~\text{m}, 110.69~\text{m})$ which is about 82.8 m away from the optimal location of UAV 1 based on the ground-truth CKMs, and that of UAV 2 converges to $\mathbf{q}^{KNN}_{2}=(-54.64~\text{m}, -149.31~\text{m})$ which is about 66.7 m away from the optimal location of UAV 2 based on the ground-truth CKMs. The sum rate converges to $R^{KNN}=10.956~\mathrm{bps/Hz}$, with the individual rates $r^{KNN}_1 = 0.267~\mathrm{bps/Hz}$ and $r^{KNN}_2 = 10.689~\mathrm{bps/Hz}$ for the two UAV users, respectively. In summary, it is observed that the proposed algorithm based on Kriging-estimated CKMs significantly outperforms that based on {\it{KNN}}-estimated CKMs and performs close to that with ground-truth CKMs, thus showing the benefit of the Kriging algorithm again.
\begin{figure*}[!htb]
    \centering
    \subfigure[CKM of GBS 1.]{
    \includegraphics[scale=0.6]{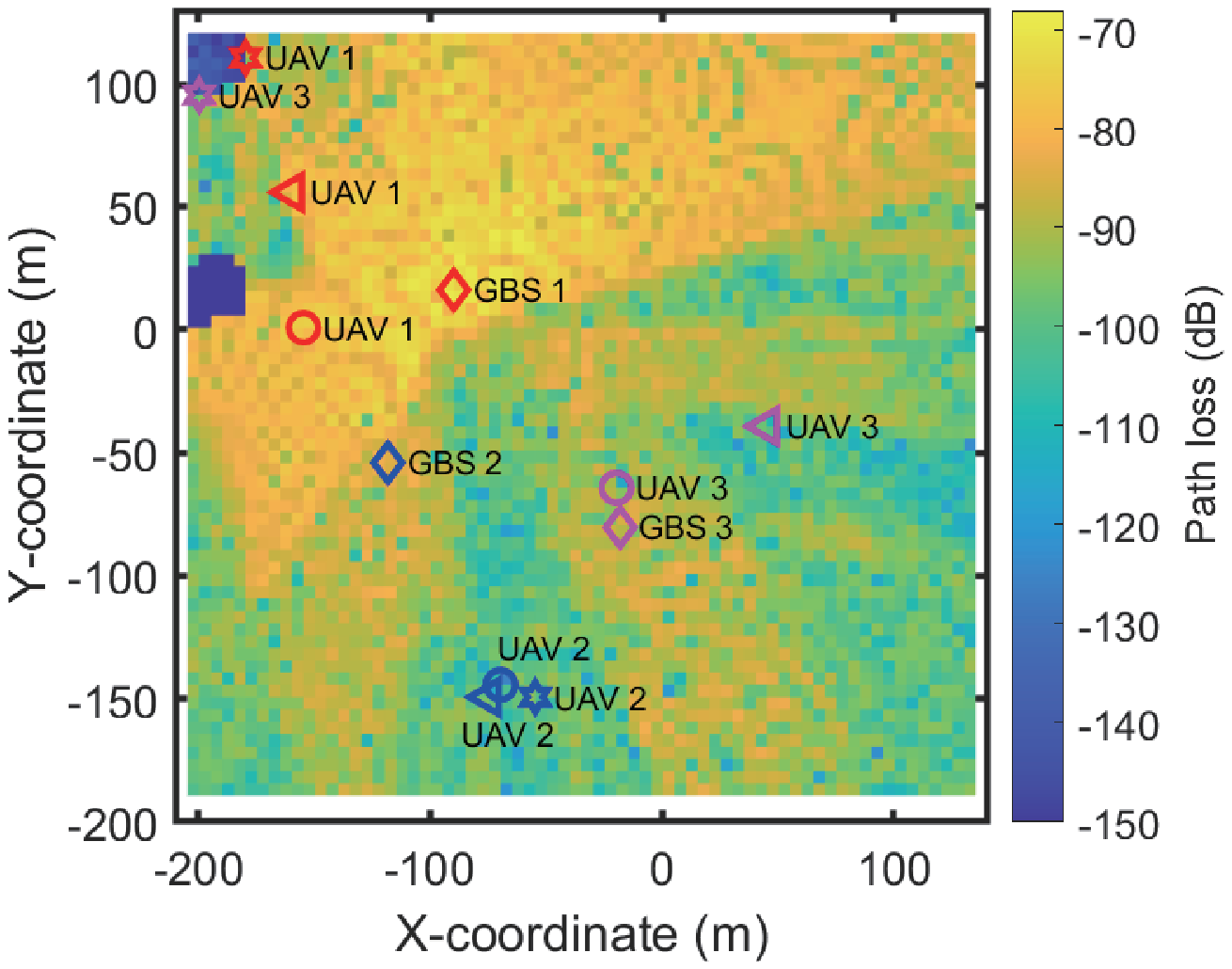}}
    \subfigure[CKM of GBS 2.]{
    \includegraphics[scale=0.6]{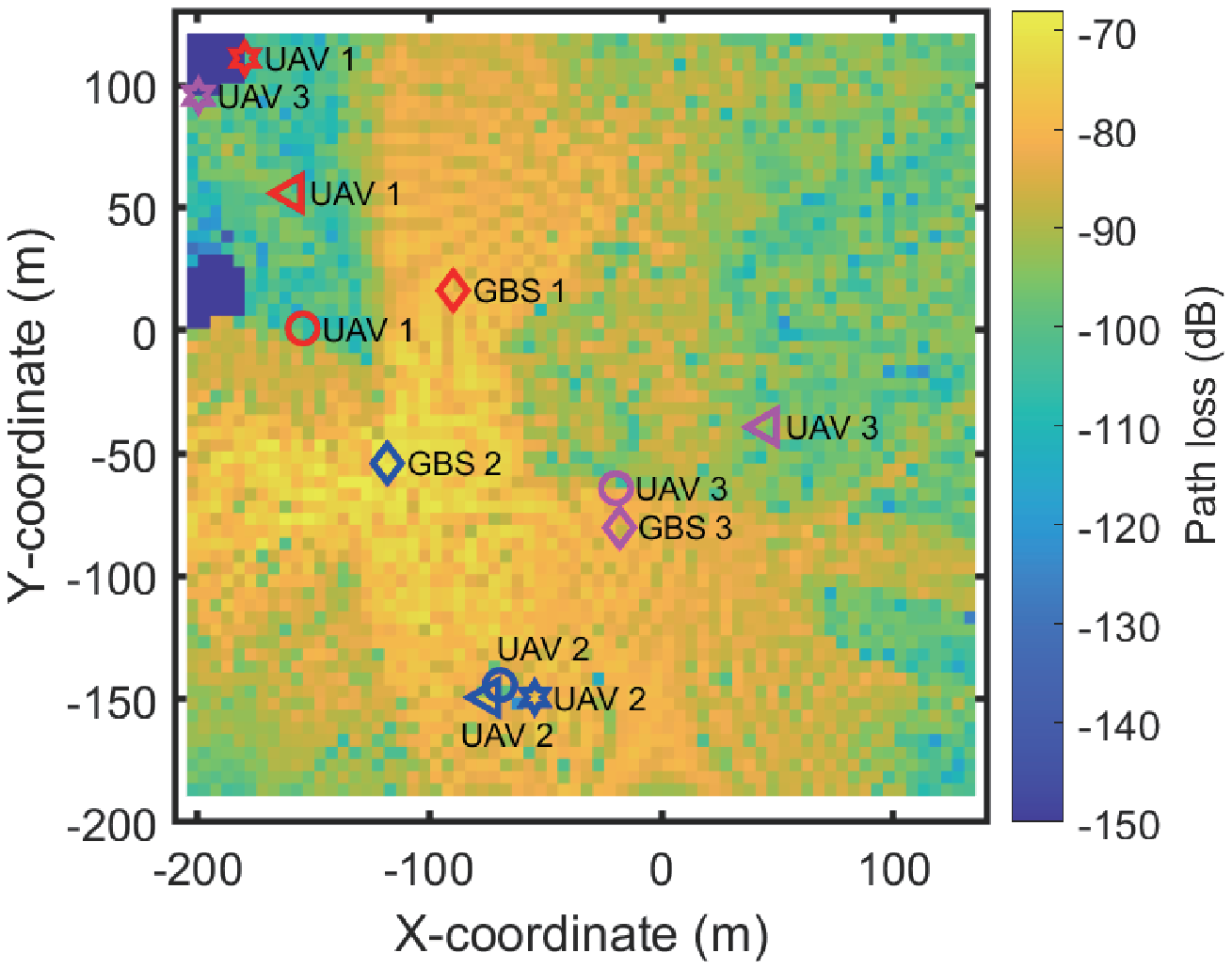}}
    \subfigure[CKM of GBS 3.]{
    \includegraphics[scale=0.6]{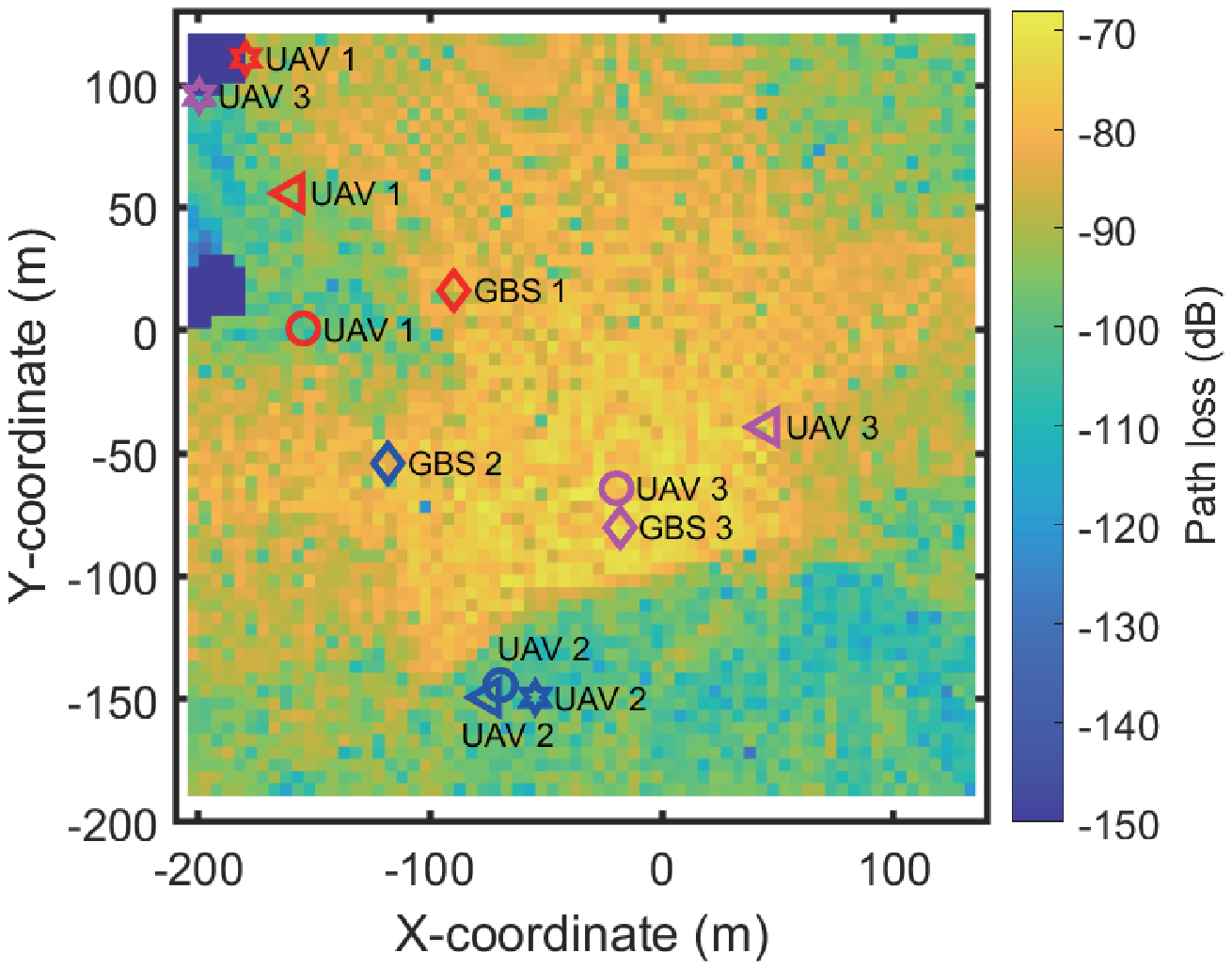}}
    \caption{Optimized UAV horizontal locations for the case with $K=3$.}
    \label{fig:CGMs-3}
\end{figure*}
\\\indent Fig. \ref{fig:CGMs-3} shows the optimized locations achieved by the proposed algorithm based on ground-truth, Kriging-estimated, and KNN-estimated CKMs for the case with $K=3$. For the proposed algorithm based on ground-truth CKMs, it is observed that the horizontal location of UAV 1 converges to $\mathbf{q}^{*}_{1}=(-154.64~\text{m}, 0.69~\text{m})$, and that of UAV 2 and UAV 3 converge to $\mathbf{q}^{*}_{2}=(-69.64~\text{m}, -144.31~\text{m})$ and $\mathbf{q}^{*}_{3}=(-19.64~\text{m}, -64.31~\text{m})$, respectively. The sum rate converges to $R=29.132~\mathrm{bps/Hz}$, with the individual rates $r_1 = 13.406~\mathrm{bps/Hz}$, $r_2 = 8.186~\mathrm{bps/Hz}$, and $r_3 = 7.540~\mathrm{bps/Hz}$.
For {\it{exhaustive search}}, the resulting sum rate is $R=30.682$ bps/Hz, which is very close to the performance achieved by the proposed algorithm. In addition, for the {\it{hovering above GBSs}}  scheme, the resulting sum rate is $R=7.114~\mathrm{bps/Hz}$, with the individual rates $r_1 = 2.766~\mathrm{bps/Hz}$, $r_2 = 3.144~\mathrm{bps/Hz}$, and $r_3 = 1.204~\mathrm{bps/Hz}$.
\\\indent It is also interesting to compare the performance of the proposed algorithm based on ground-truth CKMs with that based on Kriging-estimated and {\it{KNN}}-estimated CKMs, as shown in Fig. \ref{fig:CGMs-3}. It is observed that for the proposed algorithm based on Kriging-estimated CKMs, the horizontal location of UAV 1 converges to $\mathbf{q}^{Kriging}_{1}=(-159.64~\text{m}, 55.69~\text{m})$ which is about 55.2 m away from the optimal location of UAV 1 based on the ground-truth CKMs, and that of UAV 2 converges to $\mathbf{q}^{Kriging}_{2}=(-74.64~\text{m}, -149.31~\text{m})$ which deviates from the optimal location of UAV 2 based on the ground-truth CKMs with a distance of about 7.1 m, and that of UAV 3 converges to $\mathbf{q}^{Kriging}_{3}=(45.3612~\text{m}, -39.3143~\text{m})$ which is about 69.6 m away from the optimal location of UAV 3 based on the ground-truth CKMs. The sum rate converges to $R^{Kriging}=24.795~\mathrm{bps/Hz}$, with the individual rates $r^{Kriging}_1 = 10.706~\mathrm{bps/Hz}$, $r^{Kriging}_2 = 7.314~\mathrm{bps/Hz}$ and $r^{Kriging}_3 = 6.775~\mathrm{bps/Hz}$. In addition, it is observed that for the proposed algorithm based on {\it{KNN}}-estimated CKMs, the horizontal location of UAV 1 converges to $\mathbf{q}^{KNN}_{1}=(-179.64~\text{m}, 110.69~\text{m})$ which is about 112.8 m away from the optimal location of UAV 1 based on the ground-truth CKMs, and that of UAV 2 converges to $\mathbf{q}^{KNN}_{2}=(-54.64~\text{m}, -149.31~\text{m})$ which is about 15.8 m away from the optimal location of UAV 3 based on the ground-truth CKMs, and that of UAV 3 converges to $\mathbf{q}^{KNN}_{3}=(-199.639~\text{m}, 95.6857~\text{m})$ which is about 272.9 m away from the optimal location of UAV 3 based on the ground-truth CKMs. The sum rate converges to $R^{KNN}=10.987~\mathrm{bps/Hz}$, with the individual rates $r^{KNN}_1 = 0.951~\mathrm{bps/Hz}$, $r^{KNN}_2 = 10.036~\mathrm{bps/Hz}$ and $r^{KNN}_3 = 0.000~\mathrm{bps/Hz}$.
\begin{figure}[ht]
    \centering
    \includegraphics[scale=0.9]{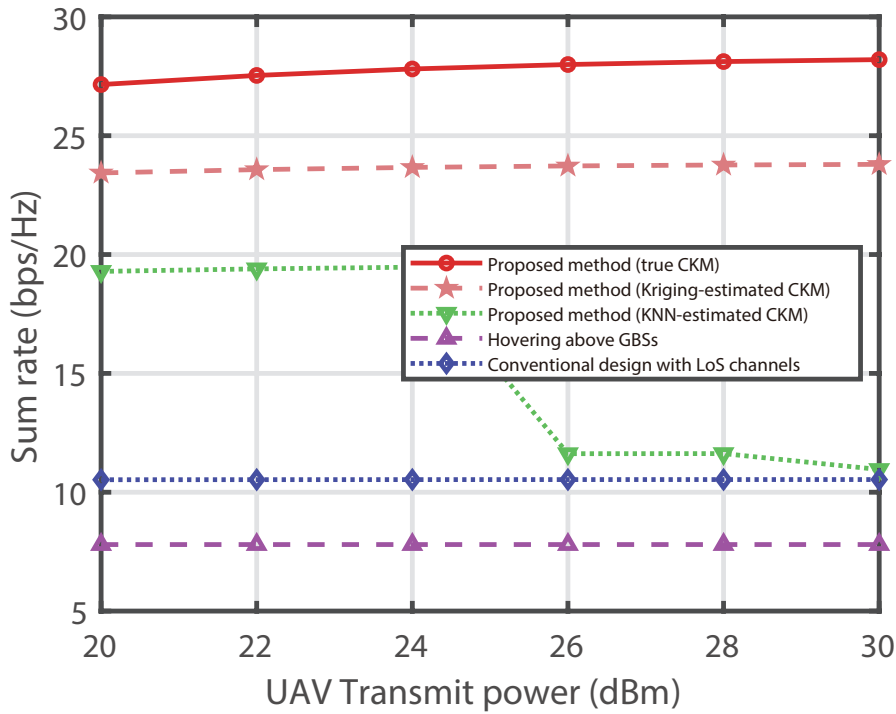}
    \caption{Performance comparison for the case with $K=2$.}
    \label{performance1}
\end{figure}\begin{figure}[ht]
    \centering
    \includegraphics[scale=0.9]{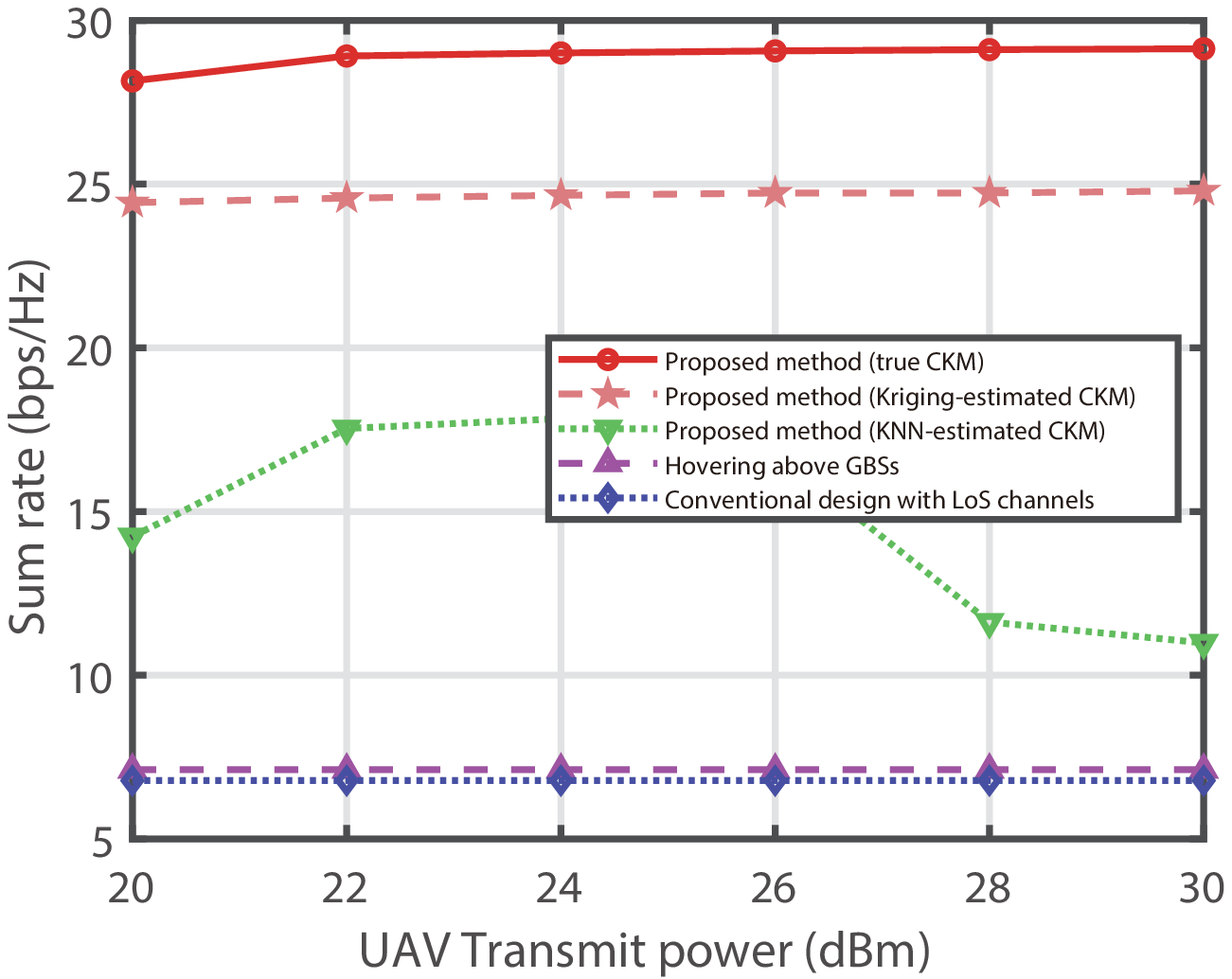}
    \caption{Performance comparison for the case with $K=3$.}
    \label{performance2}
\end{figure}
\\\indent Figs. \ref{performance1} and \ref{performance2} show the sum rate versus the UAV transmit power for the cases with $K=2$ and $K=3$, respectively. It is observed that as the UAV transmit power increases, the sum rate of all UAVs achieved by the proposed algorithm is much higher than that achieved by the {\it{hovering above GBSs}} scheme. Furthermore, compared with our proposed algorithm, the {\it{conventional design with LoS channels}} scheme is observed to achieve poor performance, as it fails to characterize the actual channel environments. Compared with the proposed algorithm based on ground-truth CKMs, it is observed that as the transmit power increases, the proposed algorithm based on Kriging-estimated CKMs achieves a sum rate of all UAV users close to that based on the ground-truth CKMs. Furthermore, for comparison, the proposed algorithm based on {\it{KNN}}-estimated CKMs is observed to achieve a poor performance as the {\it{KNN}}-estimated CKMs suffer higher estimation errors compared with the Kriging-estimated CKMs. This demonstrates that when only a small number of sampled measurements are available, the proposed algorithm based on the Kriging-estimated CKMs can achieve a close performance to that based on ground-truth CKMs with a slight sacrifice for the sum rate of about $4-5~\mathrm{bps/Hz}$.
\section{Conclusion}
\label{sec5}
This paper investigated the problem of UAV-enabled wireless networks with CKMs. First, we study the construction of sophisticated CKMs based on the actual wireless communication environments when there is only a limited number of measurement points available. Towards this end, we exploited a data-driven interpolation technique, namely the Kriging method, to construct the CKMs based on a linear interpolation model, for which the weighting factors are obtained by minimizing the variance of the estimation error between the sampled measurement points and the corresponding estimation. The numerical results showed the Kriging method achieves a CKM that is close to the ground-truth, and outperforms the KNN and the simplified LoS path loss channel model methods, in terms of the lower mean absolute error. Second, we considered a multi-UAV wireless network with CKMs, in which we maximized the weighted sum rate by jointly optimizing the UAV placement locations. Note that the CKMs are site-specific discrete databases without any closed-form expression for the channel gains, and thus the considered objective function is non-differentiable and cannot be solved by conventional convex or non-convex optimization techniques. To tackle this challenge, we proposed a novel iterative algorithm based on derivative-free optimization to obtain a high-quality solution, the key idea of which is to iteratively construct a quadratic function to approximate the objective function. Numerical results showed that our proposed algorithm achieves a close performance to exhaustive search but with much lower implementation complexity, and also outperforms other benchmark schemes. It was also shown that the proposed algorithm with Kriging-estimated CKMs based on a small number of measurement points achieves a performance close to that with ground-truth CKMs and outperforms that with KNN-estimated CKMs.


\begin{thebibliography}{1}
\bibliographystyle{IEEEbib}
\bibitem{conf}
H. Li, P. Li, J. Xu, J. Chen, and Y. Zeng, ``Derivative-free placement optimization for multi-UAV wireless networks with channel knowledge map,'' in {\it{Proc. IEEE ICC Workshop}}, 2022.

\bibitem{UAV0}
Y. Zeng, Q. Wu, and R. Zhang, ``Accessing from the sky: A tutorial on UAV communications for 5G and beyond,'' {\it{Proc. IEEE}}, vol. 107, no. 12, pp. 2327--2375, Dec. 2019.

\bibitem{UAVtuto}
Y. Zeng, J. Lyu, and R. Zhang, ``Cellular-connected UAV: Potentials, challenges and promising technologies,'' {\it{IEEE Wireless Commun.}}, vol. 26, no. 1, pp. 120--127, Feb. 2019.
\bibitem{UAV3}
Q. Wu, J. Xu, Y. Zeng, D. W. K. Ng, N. Al-Dhahir, R. Schober, and A. L. Swindlehurst, ``A comprehensive overview on 5G-and-beyond networks with UAVs: From communications to sensing and intelligence,'' {\it{IEEE J. Sel. Areas Commun.}}, vol. 39, no. 10, pp. 2912--2945, Oct. 2021.
\bibitem{UAV6}
M. Mozaffari, W. Saad, M. Bennis, Y. -H. Nam, and M. Debbah, ``A tutorial on UAVs for wireless networks: Applications, challenges, and open problems,'' {\it{IEEE Commun. Surveys Tuts.}}, vol. 21, no. 3, pp. 2334--2360, Third Quarter 2019.
\bibitem{ZYInitial}
Y. Zeng, R. Zhang, and T. J. Lim, ``Throughput maximization for UAV-enabled mobile relaying systems,'' {\it{IEEE Trans. Commun.}}, vol. 64, no. 12, pp. 4983--4996, Dec. 2016.
\bibitem{Chen1}
J. Chen, U. Mitra, D. Gesbert, ``3D urban UAV relay placement: linear complexity algorithm and analysis,'' {\it{IEEE Trans. Wireless Commun.}}, vol. 20, no. 8, pp. 5243–-5257, Mar. 2021. 
\bibitem{Chen2}
J. Chen and D. Gesbert, ``Efficient local map search algorithms for the placement of flying relays,'' {\it{IEEE Trans. Wireless Commun.}}, vol. 19, no. 2, pp. 1305--1319, Feb. 2020.
\bibitem{QQ}
Q. Wu, J. Xu, and R. Zhang, ``Capacity characterization of UAV-enabled two-user broadcast channel,'' {\it{IEEE J. Sel. Areas Commun.}}, vol. 36, no. 9, pp. 1955--1971, Sep. 2018.
\bibitem{lipeiming2}
P. Li and J. Xu, ``Fundamental rate limits of UAV-enabled multiple access channel with trajectory optimization,'' {\it{IEEE Trans. Wireless Commun.}}, vol. 19, no. 1, pp. 458--474, Jan. 2020.
\bibitem{CYou2}
C. You and R. Zhang, ``3D trajectory optimization in rician fading for UAV-enabled data harvesting,'' {\it{IEEE Trans. Wireless Commun.}}, vol. 18, no. 6, pp. 3192--3207, Jun. 2019.
\bibitem{lipeiming}
P. Li and J. Xu, ``Placement optimization for UAV-enabled wireless networks with multi-hop backhauls,'' {\it{J. Commun. Inf. Netw.}}, vol. 3, no. 4, pp. 64--73, Dec. 2018.
\bibitem{lixinmin}
X. Li and J. Xu, ``Positioning optimization for sum-rate maximization in UAV-enabled interference channel,'' {\it{IEEE Signal Process. Lett.}}, vol. 26, no. 10, pp. 1466--1470, Oct. 2019.
\bibitem{Valiulahi}
I. Valiulahi and C. Masouros, ``Multi-UAV deployment for throughput maximization in the presence of co-channel interference,'' {\it{IEEE Internet Things J.}}, vol. 8, no. 5, pp. 3605--3618, Mar. 2021.
\bibitem{QQ2}
Q. Wu, Y. Zeng, and R. Zhang, ``Joint trajectory and communication design for multi-UAV enabled wireless networks,'' {\it{IEEE Trans. Wireless Commun.}}, vol. 17, no. 3, pp. 2109--2121, Mar. 2018.
\bibitem{CKM}
Y. Zeng and X. Xu, ``Toward environment-aware 6G communications via channel knowledge map,'' {\it{IEEE Wireless Commun.}}, vol. 28, no. 3, pp. 84--91, Jun. 2021
\bibitem{R1} D. Wu, Y. Zeng, S. Jin, and R. Zhang, ``Environment-aware and training-free beam alignment for mmWave massive MIMO via channel knowledge map,''  in {\it{Proc. IEEE ICC Workshops}}, Jun. 2021, pp. 1--7.
\bibitem{R2} D. Ding, D. Wu, Y. Zeng, S. Jin, and R. Zhang, ``Environment-aware beam selection for IRS-aided communication with channel knowledge map,''  in {\it{Proc. IEEE Globecom Workshops}}, Dec. 2021, pp. 1--6.
\bibitem{SZhang}
S. Zhang and R. Zhang, ``Radio map-based 3D path planning for cellular-connected UAV,'' {\it{IEEE Trans. Wireless Commun.}}, vol. 20, no. 3, pp. 1975--1989, Mar. 2021.
\bibitem{SBi}
S. Bi, J. Lyu, Z. Ding, and R. Zhang, ``Engineering radio maps for wireless resource management,'' {\it{IEEE Wireless Commun.}}, vol. 26, no. 2, pp. 133--141, Apr. 2019.
\bibitem{RLevie}
R. Levie, \c C. Yapar, G. Kutyniok, and G. Caire, ``RadioUNet: Fast radio map estimation with convolutional neural networks,'' {\it{IEEE Trans. Commun.}}, vol. 20, no. 6, pp. 4001--4015, Jun. 2021.
\bibitem{LWJ}
W. Liu and J. Chen, ``UAV-aided radio map construction for wireless communications and localization,'' 2021. [Online] Available: {\url{https://arxiv.org/abs/2107.10574}}
\bibitem{mo}
X. Mo, Y. Huang, and J. Xu, ``Radio-map-based robust positioning optimization for UAV-enabled wireless power transfer,'' {\it{IEEE Wireless Commun. Lett.}}, vol. 9, no. 2, pp. 179--183, Feb. 2020.
\bibitem{YHuang}
Y. Huang, X. Mo, J. Xu, L. Qiu, and Y. Zeng, ``Online maneuver design for UAV-enabled NOMA systems via reinforcement learning,'' in {\it{Proc. IEEE WCNC}}, Apr. 2020, pp. 1--6.
\bibitem{ZY2}
Y. Zeng, X. Xu, S. Jin, and R. Zhang, ``Simultaneous navigation and radio mapping for cellular-connected UAV with deep reinforcement learning,'' {\it{IEEE Trans. Wireless Commun.}}, vol. 20, no. 7, pp. 4205--4220, Jul. 2021.
\bibitem{juntingc}
J. Chen, U. Mitra, and D. Gesbert, ``3D urban UAV relay placement: Linear complexity algorithm and analysis,'' {\it{IEEE Trans. Wireless Commun.}}, vol. 20, no. 8, pp. 5243--5257, Aug. 2021.
\bibitem{CYou}
C. You and R. Zhang, ``Hybrid offline-online design for UAV-enabled data harvesting in probabilistic LoS channels,'' {\it{ IEEE Trans. Wireless Commun.}}, vol. 19, no. 6, pp. 3753--3768, Jun. 2020.
\bibitem{AAi}
A. Al-Hourani, S. Kandeepan, and A. Jamalipour, ``Modeling air-to-ground path loss for low altitude platforms in urban environments,'' in {\it{Proc. IEEE GLOBECOM}}, Dec. 2014, pp. 1--6.
\bibitem{MMo}
M. Mozaffari, W. Saad, M. Bennis, and M. Debbah, ``Optimal transport theory for power-efficient deployment of unmanned aerial vehicles,'' in {\it{Proc. IEEE ICC}}, May 2016, pp. 1--6.
\bibitem{KRizk}
K. Rizk, J. -. Wagen, and F. Gardiol, ``Two-dimensional ray-tracing modeling for propagation prediction in microcellular environments,'' {\it{IEEE Trans. Veh. Technol.}}, vol. 46, no. 2, pp. 508--518, May 1997.
\bibitem{NSuga}
N. Suga, R. Sasaki, M. Osawa, and T. Furukawa, ``Ray tracing acceleration using total variation norm minimization for radio map simulation,'' {\it{ IEEE Wireless Commun. Lett.}}, vol. 10, no. 3, pp. 522--526, Mar. 2021.
\bibitem{EM}
K. Li, P. Li, Y. Zeng, and J. Xu, ``Channel knowledge map for environment-aware communications: EM algorithm for map construction,'' in {\it{Proc. IEEE WCNC}}, May 2022, pp. 1--7.
\bibitem{MLE}
J. Chen, U. Yatnalli, and D. Gesbert, ``Learning radio maps for UAV-aided wireless networks: A segmented regression approach,'' in {\it{Proc. IEEE ICC}}, May 2017, pp. 1--6.
\bibitem{DGP}
X. Wang, X. Wang, S. Mao, J. Zhang, S. C. G. Periaswamy, and J. Patton, ``Indoor radio map construction and localization with deep gaussian processes,'' {\it{IEEE Trans. Veh. Technol.}}, vol. 7, no. 11, pp. 11238--11249, Nov. 2020.
\bibitem{UMasood}
U. Masood, H. Farooq, and A. Imran, ``A machine learning based 3D propagation model for intelligent future cellular networks,'' in {\it{Proc. IEEE GLOBECOM}}, Dec. 2019, pp. 1--6.
\bibitem{YTeganya}
Y. Teganya and D. Romero, ``Deep completion autoencoders for radio map estimation,'' {\it{IEEE Trans. Wireless Commun.}}, vol. 21, no. 3, pp. 1710--1724, Mar. 2022.
\bibitem{KSuto}
K. Suto, S. Bannai, K. Sato, K. Inage, K. Adachi, and T. Fujii, ``Image-driven spatial interpolation with deep learning for radio map construction,'' {\it{IEEE Wireless Commun. Lett.}}, vol. 10, no. 6, pp. 1222--1226, Jun. 2021.
\bibitem{IDW}
D. Denkovski, V. Atanasovski, L. Gavrilovska, J. Riihij\"arvi, and P. M\"ah\"onen, ``Reliability of a radio environment map: Case of spatial interpolation techniques,'' in {\it{Proc. CROWNCOM}}, Jun. 2012, pp. 248-253.
\bibitem{knn}
R. Deng, Z. Jiang, S. Zhou, S. Cui, and Z. Niu, ``A two-step learning and interpolation method for location-based channel database construction,'' in {\it{Proc. IEEE GLOBECOM}}, Dec. 2018, pp. 1-6.
\bibitem{kriging3}
K. Sato and T. Fujii, ``Kriging-based interference power constraint: Integrated design of the radio environment map and transmission power,'' {\it{IEEE Trans. Cogn. Commun.
Netw.}}, vol. 3, no. 1, pp. 13--25, Mar. 2017.
\bibitem{toolbook}
J. Nocedal and S. J. Wright, {\it{Numerical optimization}}, Springer, 2006.
\bibitem{N. Cressie}
N. Cressie, ``Spatial prediction and ordinary kriging,'' {\it{Math. Geol.}}, vol. 20, no. 4, pp. 405--421, May 1988.
\bibitem{marazzi2002wedge}
M.~Marazzi and J.~Nocedal, ``Wedge trust region methods for derivative free optimization,'' \emph{Mathematical Programming}, vol.~91, no.~2, pp.
  289--305, 2002.
\bibitem{toolbox}
Lindon Roberts, {\it{Trustregion: Trust-region subproblem solver}}, 2021. [Online] Available: {\url{https://github.com/lindonroberts/trust-region}}
\bibitem{BOBYQA}
Powell, Michael JD, ``The BOBYQA algorithm for bound constrained optimization without derivatives,'' {\it{Cambridge NA Report}}, NA2009/06, 26--46.
\bibitem{WI}
Remcom, {\it{Wireless Insite}}. [Online] Available: {\url{https://www.remcom.com/wireless-insite-em-propagation-software}}
\end{thebibliography}
\end{document}